\documentclass[fleqn,usenatbib]{mnras}

\usepackage{newtxtext,newtxmath}
\usepackage[T1]{fontenc}

\DeclareRobustCommand{\VAN}[3]{#2}
\let\VANthebibliography\thebibliography
\def\thebibliography{\DeclareRobustCommand{\VAN}[3]{##3}\VANthebibliography}

\usepackage{xspace}
\usepackage{graphicx}
\usepackage{natbib}
\usepackage{longtable}
\usepackage{booktabs}
\usepackage{array}
\usepackage{pdflscape}
\usepackage{orcidlink}
\newcommand\tess{\textit{TESS}\xspace}
\newcommand\exofast{\texttt{EXOFASTv2}\xspace}
\newcommand\mj{M$_{\rm J}$\xspace}
\newcommand\msun{M$_\odot$\xspace}

\newcommand\rj{R$_{\rm J}$\xspace}

\newcommand{\bjdtdb}{\ensuremath{\rm {BJD_{TDB}}}}
\newcommand{\msol}{M$_\odot$\xspace}
\newcommand{\teff}{T$_{\rm eff}$\xspace}

\makeatletter
\newcommand\tablesize{\@setfontsize\tablesize{7.5}{7.5}}
\makeatother

\newcommand{\confirmedtoi}{TOI-4138\xspace}

\newcommand{\totHJs}{640} % controversial flag = 0, period < 10, mass = [0.25, 13] + 4 (my paper)
\newcommand{\transitingHJs}{588} % controversial flag = 0, period < 10, mass = [0.25, 13], discovery method = 'transit', radius = 'not null' + 4 (my paper)
\newcommand{\selfconsistentHJs}{83}
\newcommand{\NEAdate}{2025 May 8}
    {\clearpage\pagebreak[4]\global\pdfpageattr\expandafter{\the\pdfpageattr/Rotate 90}}%
    {\clearpage\pagebreak[4]\global\pdfpageattr\expandafter{\the\pdfpageattr/Rotate 0}}%

\graphicspath{{./}{figures/}}

\title[MEEP II: Super-Jupiters and Li-rich Hosts]{Migration and Evolution of giant ExoPlanets (MEEP) II: Super-Jupiters and Lithium-rich Host Stars}

\author[Schulte et al.]{Jack Schulte$^{1\orcidlink{0000-0002-7382-0160}}$, % confirmed
Joseph E. Rodriguez$^{1\orcidlink{0000-0001-8812-0565}}$, % confirmed
David~W.~Latham$^{2\orcidlink{0000-0001-9911-7388}}$, % confirmed
Joshua V. Shields$^{3\orcidlink{0000-0002-1560-5286}}$, % confirmed
Noah Vowell$^{1\orcidlink{0000-0002-0701-4005}}$, % confirmed
\newauthor
Melinda Soares-Furtado$^{4,5\orcidlink{0000-0001-7493-7419}}$, % confirmed
Brooke Kotten$^{4,6\orcidlink{0009-0008-5864-9415}}$, % confirmed
Xian-Yu Wang$^{7\orcidlink{0000-0002-0376-6365}}$, % confirmed
Karen A.~Collins$^{2\orcidlink{0000-0001-6588-9574}}$, % confirmed
Allyson Bieryla$^{2\orcidlink{0000-0001-6637-5401}}$, % confirmed
\newauthor
Samuel~N.~Quinn$^{2\orcidlink{0000-0002-8964-8377}}$,
Paul Benni$^{8\orcidlink{0000-0001-6981-8722}}$, % confirmed
Catherine A. Clark$^{9\orcidlink{0000-0002-2361-5812}}$, % confirmed
Matthew W. Craig$^{10\orcidlink{0000-0001-7988-8919}}$, % confirmed
Mara L. DeRung$^{10}$, % confirmed
\newauthor
Jason~D.~Eastman$^{2\orcidlink{0000-0003-3773-5142}}$, % confirmed
Zahra~Essack$^{11\orcidlink{0000-0002-2482-0180}}$, % confirmed
Phil Evans$^{12\orcidlink{0000-0002-5674-2404}}$, % confirmed
Rebecca Gore$^{13,14}$, % confirmed
Steve B. Howell$^{14\orcidlink{0000-0002-2532-2853}}$,
\newauthor
John F.~Kielkopf$^{15\orcidlink{0000-0003-0497-2651}}$, % confirmed
Colin Littlefield$^{15,16}$, % Confirmed 
Andrew W. Mann$^{16\orcidlink{0000-0003-3654-1602}}$, % confirmed
Giuseppe Marino$^{17\orcidlink{0000-0001-8134-0389}}$, % Confirmed 
Don J. Radford$^{18\orcidlink{0000-0002-3940-2360}}$, % confirmed
\newauthor
Chris Stockdale$^{19\orcidlink{0000-0003-2163-1437}}$, % confirmed
Ivan A. Strakhov$^{20\orcidlink{0000-0003-0647-6133}}$, % confirmed
Thiam-Guan Tan$^{21\orcidlink{0000-0001-5603-6895}}$, % confirmed
Michael~Vezie$^{5}$, % Confirmed 
Songhu Wang$^{7\orcidlink{0000-0002-7846-6981}}$, % confirmed
\newauthor
Emily Watson$^{10}$, % confirmed
Samuel W. Yee$^{2\orcidlink{0000-0001-7961-3907}}$, % confirmed
and Carl Ziegler$^{22\orcidlink{0000-0002-0619-7639}}$ % confirmed
\\
$^{1}$Center for Data Intensive and Time Domain Astronomy, Department of Physics and Astronomy, Michigan State University, East Lansing, MI 48824, USA\\
$^{2}$Center for Astrophysics | Harvard \& Smithsonian, 60 Garden St, Cambridge, MA 02138, USA\\
$^{3}$Department of Physics and Astronomy, Michigan State University, East Lansing, MI 48824, USA\\
$^{4}$Department of Astronomy, University of Wisconsin–Madison, 475 N Charter St., Madison, WI 53706, USA\\
$^{5}$Department of Physics and Kavli Institute for Astrophysics and Space Research, Massachusetts Institute of Technology, Cambridge, MA 02139, USA\\
$^{6}$Department of Astronomy, University of Michigan, Ann Arbor, MI 48109, USA\\
$^{7}$Department of Astronomy, Indiana University, Bloomington, IN 47405, USA\\
$^{8}$Acton Sky Portal (private observatory), Acton, MA, USA\\
$^{9}$NASA Exoplanet Science Institute, IPAC, California Institute of Technology, Pasadena, CA 91125, USA\\
$^{10}$Department of Physics and Astronomy, Minnesota State University Moorhead, 1104 7th Avenue South, Moorhead, MN 56563, USA\\
$^{11}$Department of Physics and Astronomy, The University of New Mexico, 210 Yale Blvd NE, Albuquerque, NM 87106, USA\\
$^{12}$El Sauce Observatory, Coquimbo Province, Chile\\
$^{13}$Bay Area Environmental Research Institute, Moffett Field, CA 94035, USA\\
$^{14}$NASA Ames Research Center, Moffett Field, CA 94035, USA\\
$^{15}$Department of Physics and Astronomy, University of Louisville, Louisville, KY 40292, USA\\
$^{16}$Department of Physics and Astronomy, The University of North Carolina at Chapel Hill, Chapel Hill, NC 27599, USA\\
$^{17}$Gruppo Astrofili Catanesi, Catania, Italy\\
$^{18}$Brierfield Observatory, New South Wales, Australia\\
$^{19}$Hazelwood Observatory, Australia\\
$^{20}$Sternberg Astronomical Institute, Lomonosov Moscow State University, Universitetsky pr. 13, Moscow 119234, Russia\\
$^{21}$Perth Exoplanet Survey Telescope, Perth, Western Australia, Australia\\
$^{22}$Department of Physics, Engineering and Astronomy, Stephen F. Austin State University, 1936 North St, Nacogdoches, TX 75962, USA
}

\date{Accepted XXX. Received YYY; in original form ZZZ}

\pubyear{\the\year{}}

\begin{document}
\label{firstpage}
\pagerange{\pageref{firstpage}--\pageref{lastpage}}
\maketitle

\begin{abstract}
Although hot Jupiters were the first exoplanets discovered orbiting main sequence stars, the dominant mechanisms through which they form and evolve are not known. To address the questions surrounding their origins, the Migration and Evolution of giant ExoPlanets (MEEP) survey aims to create a complete, magnitude-limited ($G<$12.5) sample of hot Jupiters that can be used to constrain the frequency of different migration pathways. NASA's Transiting Exoplanet Survey Satellite provides the unique combination of sky-coverage and photometric precision to achieve this goal, which will likely be a key result of the mission. In this second installment of the MEEP survey, we reanalyze one benchmark hot Jupiter system, \confirmedtoi, and discover four additional super-Jupiters which are each more than five times as massive as Jupiter: TOI-4773 b, TOI-5261 b, TOI-5350 b, and TOI-6420 b. One of these planets, TOI-5261 b, is 11.49 times the mass of Jupiter, nearly massive enough to ignite deuterium fusion, and has an eccentric ($e = 0.1585$) orbit. \confirmedtoi, TOI-4773, TOI-5350, and TOI-6420 each have lithium absorption features in their spectra. \confirmedtoi is an F-type subgiant with a lithium equivalent width of $120. \pm 13$ m\AA, which is $\sim 4.5\sigma$ larger than the median lithium equivalent width of a control sample of 1381 similar stars, making \confirmedtoi a compelling candidate for planetary engulfment.
\end{abstract}

\begin{keywords}
planets and satellites: detection -- planets and satellites: gaseous planets -- stars: abundances
\end{keywords}

\renewcommand{\arraystretch}{1.6} % to make tables more readable

%%%%%%%%%%%%%%%%%%%%%%%%%%%%%%%%%%%%%%%%%%%%%%%%%%

%%%%%%%%%%%%%%%%% BODY OF PAPER %%%%%%%%%%%%%%%%%%

\section{Introduction}

In the three decades since the first discovery of a short-period gas giant, or hot Jupiter (HJ), orbiting a main sequence star \citep{Mayor:1995}, \totHJs{} HJs\footnote{Retrieved from \url{https://exoplanetarchive.ipac.caltech.edu/} on \NEAdate} have been confirmed by a variety of ground-based and space-based facilities. These discoveries have led to tentative trends, correlations, and occurrence rates \citep[e.g.,][]{Huang:2016, Bryan:2016, Bonomo:2017, Ikwut-Ukwa:2022, Rodriguez:2023, Yee:2023b, Zink:2023} and a host of unanswered questions. While the constructed population is large, it suffers from a lack of self-consistency as a consequence of several factors important to the discovery of these systems: (1) the discoveries have spanned a large time period, over which instrument sensitivities and analysis techniques have evolved, (2) the discoveries were made using many different instruments and fitting software that operate under different assumptions, and (3) the decisions made during the fitting process vary by author, leading to significantly different reported parameters, especially the planet's orbital eccentricity. To address these issues, we introduced the Migration and Evolution of giant ExoPlanets (MEEP) survey in \cite{Schulte:2024}, which is aimed at constructing a complete, self-consistent, sample of HJs orbiting FGK stars brighter than a \textit{Gaia} \textit{G}-band magnitude limit of 12.5. Our survey, combined with other collaborating efforts \citep{Rodriguez:2021, Ikwut-Ukwa:2022, Yee:2022, Yee:2023a, Rodriguez:2023, Yee:2025}, aims to construct this sample in the coming years using space-based photometry from the Transiting Exoplanet Survey Satellite (\tess; \citealt{Ricker:2015}) and the open-source global fitting software \exofast\footnote{\url{https://github.com/jdeast/EXOFASTv2}} \citep{Eastman:2019}. The use of consistent priors, \tess data, and \exofast ensures self-consistency and the extensive follow-up efforts of the \tess Follow-up Observing Program (TFOP; \citealt{Collins:2018}) enables the efficient construction of this sample.

While there is much that can be done with a large, self-consistent catalog of HJs, one of the primary goals of this survey is to assess the existing theories explaining the evolutionary pathways of HJs. These theories can be grouped into three categories: in-situ formation \citep{Batygin:2016}, gas-disk migration \citep{Goldreich:1980, Lin:1986}, and high-eccentricity tidal migration \citep[e.g.,][]{Kozai:1962, Lidov:1962, Rasio:1996, Naoz:2016}. While in-situ formation has been proposed as the most likely mechanism for some individual systems \citep[e.g.,][]{Poon:2021}, it has been argued that it is unlikely to be the dominant mechanism for HJ formation due to the implausibility of a rapid build-up of planetary material in regions of the disk where feeding zones are small \citep{Lee:2014, Dawson:2018}. Gas-disk migration and high-eccentricity tidal migration, the two flavors of ex-situ HJ formation, are both plausible for most of the observed HJ systems. Migration through a gas-rich circumstellar disk is possible without scattering nearby planets and is expected to result in multi-planet systems with low eccentricity. Several systems that exhibit these traits have been discovered \citep[e.g.,][]{Becker:2015, Canas:2019, Wu:2023}, but it appears to be rare for HJs to have neighboring planets \citep{Huang:2016}. On the other hand, high-eccentricity migration occurs when a planet exchanges angular momentum with another planet or star, resulting in a highly eccentric, and possibly misaligned, orbit, which is circularized and realigned on timescales governed by the tidal quality factor of the planet and host star \citep{Lai:2012}. This process scatters nearby planets, leading to isolated HJ systems. Depending on the tidal recircularization and realignment timescales, the planet may also leave traces of the event that led to its migration. Many isolated HJ systems have been discovered \citep{Huang:2016, Hord:2021} and a minority of these HJs have eccentric and misaligned orbits \citep{Albrecht:2012, Schulte:2024} which are more readily explained by high-eccentricity tidal migration than gas-disk migration or in-situ formation. The HJ catalog constructed by the MEEP survey aims to utilize the statistics of a large number of HJs with a well-defined selection function to determine which of these mechanisms is responsible for most HJ systems, place constraints on the frequency of each migration pathway, and uncover additional mysteries surrounding the evolution of HJs.

In this second paper of the MEEP survey, we present four discoveries of HJs and one reanalysis of a previously confirmed HJ system with a newly discovered lithium feature in the host star. Four of the five host stars in this sample have significant absorption features of the lithium doublet (Li I) at a wavelength of 6707.8 \AA. The presence of large quantities of Li in a star can be surprising as its most common stable isotope, $^7$Li, is destroyed by proton fusion at temperatures greater than $\sim 3 \times 10^6$ K \citep{Bodenheimer:1965}. Because of this, Li abundance is expected to decline as a star ages and its Li is mixed into its hotter interior, leaving most convective main sequence stars with trace amounts of Li. However, Li has been found to be overabundant in a minority of stars \citep{Chen:2001} and some of the possible explanations are enrichment by classical nova outbursts \citep{Starrfield:1978}, ingestion of planetary material \citep{Soares-Furtado:2021, Behmard:2023}, or self-enrichment by the Cameron-Fowler conveyor in red giant stars \citep{Cameron:1971}. Finally, Li has been used as a tracer for age because, barring enrichment from the aforementioned sources, the Li abundance in a pre-main sequence star decreases as the star ages \citep{Skumanich:1972, Jeffries:2023}.

% Young HJ systems can provide excellent constraints to models of planetary formation and migration. If a HJ system is confirmed to be young, its characterization becomes vital to the understanding of the early stages of planetary system evolution and the timescales that govern these processes. There are several programs (e.g., \citealt{Bouma:2019, Bouma:2020}; \citealt{Tran:2021, Tran:2024}) that aim to discover and characterize young giant planets for this purpose. The most reliable method to ascertain the age of a star, and by extension, its planetary system, is through cluster membership. In the case that a star does not appear to be associated with a cluster of stars, less precise age-dating methods can be employed, such as isochrone fitting (e.g., using \citealt{Choi:2016}), Li depletion \citep{Skumanich:1972, Jeffries:2023}, gyrochronology \citep{Bouma:2023}, and chromospheric activity indicators such as X-ray emission \citep{Jackson:2012} [MORE CITATIONS?] and calcium line emission \citep{Mamajek:2008, Zerjal:2017}. In this article, we use several of these techniques to constrain the ages of the host stars in this sample.

This article presents the reanalysis of one low-mass, low-density hot Jupiter orbiting a subgiant star with an anomalously large lithium abundance, and the confirmation of four massive super-Jupiters orbiting FGK stars. The system which we are reanalyzing, TIC 257060897, hereafter referred to by its \tess Object of Interest identifier, \confirmedtoi, was first discovered in 2022 in the \tess full frame images \citep{Montalto:2022}. In the discovery paper, its density was identified as $0.25 \pm 0.02 \mathrm{~g~cm}^{-3}$, making the planet one of the least dense known. The authors noted that its inflation is likely due to the host star's quickly increasing luminosity as it evolves off of the main sequence. The other four systems (TOI-4773, TOI-5261, TOI-5350, and TOI-6420) occupy a region in the mass space of transiting planets ($> 5$ \mj) that has a relative scarcity of objects. It appears that such massive planets are less likely, but still possible \citep{Bodenheimer:2013}, outcomes of core accretion \citep{Pollack:1996}.

In \S \ref{sec:observations} of this article, we describe the photometric and spectroscopic observations used to characterize each planet and host star and the high-resolution imaging used to rule out blended stellar companions. In \S \ref{sec:analysis}, we walk through the global fits used to determine the properties of each planetary system as well as the characterization of Li features in four of the host stars' spectra. Finally, in \S \ref{sec:discussion}, we discuss each planetary system in greater detail and place them into the context of the growing self-consistent sample of HJs being generated by the MEEP survey.

\section{Observations}\label{sec:observations}

Many different observations are required in order to confirm a transiting planet candidate as a bonafide exoplanet. In this manuscript, we used a combination of space-based and ground-based photometry, spectroscopy, and high-resolution imaging to ensure that the transit signal of each planet is on target, shows no signs of chromaticity, and that the spectra are not composite. In this section, we briefly review the observations used to confirm each planet and constrain their parameters. More details on these observations can be found in the first paper in the MEEP series \citep{Schulte:2024}.

\subsection{\tess Photometry}\label{subsec:TESS}

Each of the five targets in this paper was first observed by the \tess spacecraft and identified as a \tess Object of Interest (TOI; \citealt{Guerrero:2021}). \confirmedtoi was first established as a Community \tess Object of Interest (CTOI) on 2021 February 8 by \cite{Olmschenk:2021}, while TOI-4773, TOI-5261, TOI-5350, and TOI-6420 were each discovered in the faint-star search \citep{Kunimoto:2022} as part of the MIT Quick-Look Pipeline (QLP; \citealt{Huang:2020a, Huang:2020b, Kunimoto:2021, tessqlp}). The data were reduced both by QLP and the \tess Science Processing Operations Center (SPOC) Pipeline \citep{Jenkins:2016, Caldwell:2020, tess2min, tessffi}. In the cases where both QLP and SPOC have reduced the same data, we use the SPOC results. Each sector of \tess data used in our fits is presented in Table \ref{tab:tess}. In this table, a distinction is made between the SPOC data from full-frame images, labeled as ``TESS-SPOC," and the 2-minute and 20-second cadence SPOC data, labeled ``SPOC."

% add more details on SPOC/QLP? I go into greater detail in the MEEP 1 paper

\begin{table}
	\centering
	\caption{Summary of Observations from TESS.}
	\label{tab:tess}
	\begin{tabular}{l l l l}
		\hline
		Target & TESS Sector & Cadence (s) & Source \\
		\hline
		\confirmedtoi & 14 & 1800 & TESS-SPOC \\
		--- & 15 & 1800 & TESS-SPOC \\
		--- & 16 & 1800 & TESS-SPOC \\
		--- & 20 & 1800 & TESS-SPOC \\
		--- & 21 & 1800 & TESS-SPOC \\
		--- & 22 & 1800 & TESS-SPOC \\
		--- & 26 & 1800 & TESS-SPOC \\
		--- & 40 & 120 & SPOC \\
		--- & 41 & 120 & SPOC \\
		--- & 47 & 120 & SPOC \\
		--- & 48 & 120 & SPOC \\
		--- & 49 & 120 & SPOC \\
		--- & 53 & 120 & SPOC \\
		--- & 56 & 20 & SPOC \\
		--- & 60 & 20 & SPOC \\
		--- & 74 & 120 & SPOC \\
		--- & 75 & 120 & SPOC \\
		TOI-4773 & 34 & 600 & TESS-SPOC \\
		--- & 61 & 120 & SPOC \\
		TOI-5261 & 14 & 1800 & QLP \\
		--- & 41 & 600 & QLP \\
		--- & 54 & 600 & TESS-SPOC \\
		--- & 55 & 600 & QLP \\
		TOI-5350 & 43 & 600 & TESS-SPOC \\
		--- & 44 & 600 & TESS-SPOC \\
		--- & 71 & 200 & QLP \\
		TOI-6420 & 7 & 1800 & QLP \\
		--- & 8 & 1800 & QLP \\
		--- & 61 & 1800 & QLP \\
		\hline
	\end{tabular}
\end{table}

We downloaded each lightcurve using a custom pipeline built around the \texttt{lightkurve}\footnote{\url{https://github.com/lightkurve/lightkurve}} package, which accesses the lightcurves via the Mikulski Archive for Space Telescopes (MAST\footnote{\url{https://archive.stsci.edu/}}). Once downloaded, we flattened the lightcurves using the \texttt{choosekeplersplinev2} function in the Python package \texttt{keplersplinev2}\footnote{\url{https://github.com/avanderburg/keplersplinev2}} \citep{Vanderburg:2014}. \texttt{choosekeplersplinev2} fits a spline to the out-of-transit data using a value for the breakpoint spacing that minimizes the Bayesian information criterion. After determining the best-fit spline, we divided the entire sector by it to remove most of the stellar variability. To reduce the computational cost of our transit fits, we then chopped each lightcurve to remove unnecessary out-of-transit data, leaving a baseline of one transit duration ($T_{14}$) on either side of each transit.

The \tess space telescope is a uniquely valuable tool for the creation of a complete, magnitude-limited sample of transiting HJs. Armed with four cameras that can cover a $24^\circ \times 96^\circ$ area of the sky ($\sim 5\%$ of the entire sky) at once, \tess has observed nearly the entire sky with a minimum baseline of 27 days. As a consequence, nearly every transiting HJ orbiting a bright ($G < 12.5$) star has been observed or will be observed by \tess. However, as a trade-off for its large observing sectors, \tess has a pixel scale of $21 \arcsec$ $\mathrm{pixel}^{-1}$, which often means that there are multiple unresolved stars in each photometric aperture, a problem which worsens in more crowded fields. It is, in part, for this reason that we obtain additional photometric observations from ground-based facilities with higher spatial resolution.

\subsection{Ground-based Follow-up Photometry}\label{subsec:followup}

Following their discoveries by \tess, notices were sent to TFOP members worldwide, enabling the efficient ground-based follow-up of each target. These observations serve several important purposes. Notably, all of the follow-up observations were taken using cameras with pixel scales of $1.52\arcsec$ pixel$^{-1}$ and smaller, which is more than a factor of 13 better resolution than \tess's cameras. This allowed us to ensure that the transit signal was on the correct source, ruling out contamination by eclipsing binaries in the photometric aperture as the source of the transit-like events. Additionally, the follow-up observations were collected using a variety of filters, different from the \tess filter. This allowed us to test for evidence of significant differences in the transit depth as a function of color, commonly referred to as chromaticity. While transiting exoplanets do naturally exhibit chromaticity if they have clear atmospheres \citep[e.g.,][]{Feinstein:2023}, eclipsing binaries typically have much more significant variations in depth for corresponding changes in wavelength \citep{Tingley:2004}. In addition to their utility in ruling out false positives, our follow-up observations extended the baseline of the \tess observations to allow for better constraints on each planet's ephemeris. A well-constrained ephemeris improves the reliability of future transit epochs and therefore enables future follow-up efforts.

In this work, we collected 12 transit observations of our targets from 11 different facilities, spanning more than 85 degrees in latitude and 260 degrees in longitude. These observations are listed in Table \ref{tab:followup} along with the relevant details of each instrument and observation. All except for the PEST observation of TOI-4773 and the Feder Observatory observation of TOI-5261 were reduced using the photometry tool AstroImageJ \citep[AIJ; ][]{Collins:2017}. A detailed description of the process used to reduce photometry in AIJ is included in \S 2.2 in \cite{Schulte:2024}.

Two of the collected follow-up lightcurves were reduced using different software. The Feder Observatory's transit observation of TOI-5261 on 2022 Dec 1 was reduced using the Python package \texttt{ccdproc} \citep{Craig:2022}. The aperture photometry was then performed using \texttt{stellarphot} \citep{Craig:2024}, a Python-based photometry tool, following the conventions used in AIJ. In the case of the observation made by the PEST observatory, a custom pipeline based on {\tt C-Munipack}\footnote{\url{http://c-munipack.sourceforge.net}} was used to calibrate the images and extract the differential photometry. These lightcurves, along with all of those reduced in AIJ, are available to download on ExoFOP-TESS\footnote{\url{https://exofop.ipac.caltech.edu/tess/}}.

\begin{table*}
\tablesize
\caption{TFOP Photometric Follow-up Observations}\label{tab:followup}
\begin{tabular}{lcccccccccc}
\hline
TIC ID &  TOI $\#$  & Telescope & Obs. Date (UTC) & Tel. Size (m) & Filter & Pix. Scale (arcsec) & Phot. Aper. (arcsec) & Exp. (sec.) & Det. Params  \\
\hline

257060897 & 4138 & GAC$^{1}$ & 2021-09-03 & 0.3 & $R$ & 1.52  & 6.1 & 300 & None  \\

 415276070 & 4773 & PEST$^{2}$ & 2022-01-27 & 0.3 & $r'$  & 0.71 & 7.1 & 120   & None  \\
 & &  El Sauce & 2022-02-28 & 0.5 & $Rc$ & 1.08 & 5.4 & 30    & Airmass  \\
 & &  Brierfield & 2022-12-23 & 0.3 & $B$ & 0.74 & 5.7 & 180   & Width\_T1  \\
 & &  Hazelwood & 2023-01-20 & 0.3 & $g'$ & 0.56 & 5.6 & 240   & Airmass  \\

402828941  & 5261 &  KeplerCam &  2022-06-15 & 1.2 & $i'$  & 0.67 & 3.4 & 18  & Airmass  \\
 & &  ASP$^{3}$ & 2022-07-31  & 0.4 & $r'$ & 1.00 & 7.0 & 30  & Airmass  \\
 & &   FO$^{4}$ & 2022-08-04  & 0.4 & $i'$ & 0.56 & 5.6 & 180 & None  \\
 & &  El Sauce & 2023-06-03  & 0.5 & $B$ & 0.45 & 8.1 & 180 & Airmass  \\

68808155   & 5350  & ULMT$^{5}$ & 2022-12-01 & 0.6 & $r'$  & 0.4 & 5.2 & 64 & Airmass  \\

143526233  & 6420 & Brierfield & 2023-12-27 & 0.4 &  $R$    & 0.74 & 3.7 & 300 & Airmass  \\
& & LCOGT-CTIO$^{6}$ & 2024-03-26 & 0.4 & $g'$  & 0.73 & 5.1 & 200 & None  \\

\hline
\end{tabular}
\begin{flushleft}
\footnotesize{\textbf{\textsc{NOTE:}}
Osservatorio GAC$^{1}$, Perth Exoplanet Survey Telescope (PEST)$^{2}$, Acton Sky Portal$^{3}$, Feder Observatory$^{4}$, University of Louisville Manner Telescope (ULMT)$^{5}$, Las Cumbres Observatory Global Telescope \citep{Brown:2013}, Cerro Tololo Inter-American Observatory in Chile (CTIO)$^{6}$. All lightcurves are available on ExoFOP. See \S D from \cite{Collins:2017} for a description of the detrending parameters.}
\end{flushleft}
\end{table*}

\subsection{Spectroscopy}

In order to rule out blended eclipsing binaries and ascertain the masses and orbital eccentricities of the planets in this article, we collected ground-based high-resolution spectroscopy using two separate facilities in order to make radial velocity (RV) measurements of each star. Our five HJ systems have large radial velocity amplitudes as a consequence of their large planetary masses and their close proximity to the host stars. This allows us to place very good constraints on both the mass and eccentricity of each HJ. Additionally, these spectra can be used to rule out bright eclipsing binaries. If the eclipsing binary is bright enough to produce the transit signal and close enough to not be detected by the high resolution imaging collected of each system, it is likely to appear as an extra set of lines in the spectra. Lastly, the full RV orbit of the planet adds an additional layer of certainty that the object producing the transit signal is of planetary size and mass. This has proven to be the case for each of the systems presented in this work, as is further elaborated in the following sections. 

We collected a total of 91 RVs of the five systems using three spectrographs: the Tillinghast Reflector Echelle Spectrograph (TRES; \citealt{gaborthesis}) installed on the Tillinghast Reflector at Fred Lawrence Whipple Observatory (FLWO) in Arizona, the CHIRON spectrograph \citep{Tokovinin:2013, Paredes:2021} installed on the SMARTS 1.5-m telescope at the Cerro Tololo Inter-American Observatory in Chile, and the NEID spectrograph \citep{Schwab:2016, Halverson:2016}, installed on the WIYN 3.5-meter telescope at the Kitt Peak National Observatory in Arizona. An example RV measurement is shown for each target and instrument in Table \ref{tab:rv}. The full table of RV measurements in machine-readable format is available in the online journal. Finally, we included the RVs collected by \cite{Montalto:2022} using the High Accuracy Radial velocity Planet Searcher for the Northern hemisphere (HARPS-N) in the fit for \confirmedtoi to check its consistency with our more recent observations.

\begin{table*}
    % \centering
    \caption{The first RV measurement of each system, per instrument used.}
    \label{tab:rv}
    \begin{tabular}{l l l r r}
        \hline
        Target & Spectrograph & BJD$_{\mathrm{TDB}}$ & RV (m\,s$^{-1}$) & $\sigma_{\mathrm{RV}}$ (m\,s$^{-1}$) \\
        \hline
        TOI-4138 & TRES (Season 1) & 2460107.744224 & 9.4 & 27.9 \\
        TOI-4138 & TRES (Season 2) & 2460429.947428 & -103.9 & 22.8 \\
        TOI-4773 & CHIRON           & 2459934.7726   & 9946.0 & 72.0 \\
        TOI-5261 & TRES             & 2459694.942525 & -448.5 & 23.8 \\
        TOI-5350 & TRES             & 2459643.671565 & -5.9   & 47.2 \\
        TOI-6420 & TRES             & 2460251.982457 & -872.9 & 29.8 \\
        \hline
    \end{tabular}
    \vspace{2mm}
    
    \begin{minipage}{0.95\linewidth}
        \footnotesize
        \textbf{Note:} The full table of RVs for each system is available in machine-readable form in the online journal.
    \end{minipage}
\end{table*}

\subsubsection{TRES Spectroscopy}\label{subsubsec:tres}

We used the TRES spectrograph to observe four of the five stars presented in this article: \confirmedtoi, TOI-5261, TOI-5350, and TOI-6420. TRES is a high-resolution, fiber-fed \'echelle spectrograph mounted on the 1.5-m Tillinghast Reflector, which has a resolving power of 44,000. We reduced the spectra following the works of \cite{Buchhave:2010} and \cite{Quinn:2012}. Then, we derived multi-order relative velocities by cross-correlating the reduced spectra with a median-filtered, combined template built from all of the spectra. RV zero point offsets were corrected using standard star observations. Finally, we used the Stellar Parameter Classification (SPC) tool \citep{Buchhave:2012} to measure the effective temperature, metallicity, surface gravity, and projected equatorial radial velocity of each star. The weighted sums of these metallicity measurements were used as priors in our global fits.

While our primary purposes of these RV observations were to measure masses and eccentricities and rule out false positives, the RVs also allowed us to investigate evidence of additional companions orbiting the HJ hosts. This evidence comes in the form of long-term RV trends which indicate that an additional massive body is orbiting the HJ host on a longer orbital period than the HJ. These long-term trends are better constrained with longer RV baselines. For TOI-5261, TOI-5350, and TOI-6420, the RV baselines were 865 days, 214 days, and 106 days, respectively. We analyzed these RV datasets with a Lomb-Scargle periodogram and found no compelling evidence of periodicity (other than from each candidate transiting planet) across these baselines.

\begin{figure}
    \centering
    \includegraphics[width=0.45\textwidth,keepaspectratio]{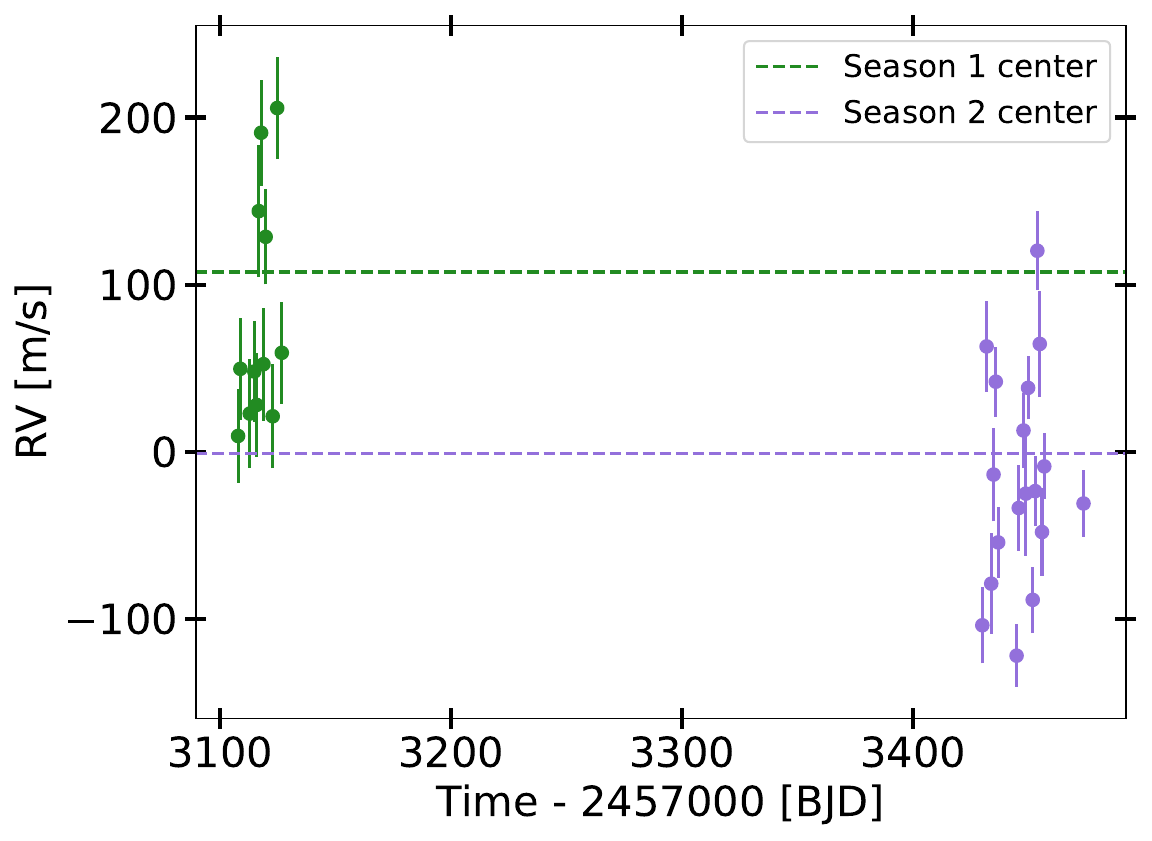}
    \caption{\confirmedtoi's RV observations from the TRES spectrograph. The first season was from UTC 12 June 2023 to UTC 01 July 2023, while the second season was from UTC 29 April 2024 to UTC 12 June 2024. The center point of the two seasons are offset by 108 m s$^{-1}$. This offset could indicate the presence of an additional, long-period companion in the system.}
    \label{fig:toi4138_2seasons}
\end{figure}

For our reanalysis of \confirmedtoi, we extended the 312-day HARPS baseline by an additional 1175 days using the TRES spectrograph. The majority of the TRES data were collected in two separate seasons: the first season spanned from UTC 12 June 2023 to UTC 01 July 2023, while the second season spanned from UTC 29 April 2024 to UTC 12 June 2024. Although these observations were taken with the same instrument, there is a clear RV offset of roughly 100 m s$^{-1}$ between the two seasons, as we illustrate in Figure \ref{fig:toi4138_2seasons}. As we did not expect a significant change in the RV zero point, this offset represents possible evidence of an additional object in the system with an orbital period longer than the observed exoplanet. Unfortunately, since we have not observed a full period of this tentative object, we cannot confidently place constraints on its properties. Therefore, we choose to treat the two seasons of TRES RVs as separate instruments, allowing the gamma, jitter, and jitter variance to vary. This enables us to focus on obtaining the best parameters for the known companion while correcting for a possible instrumental or physical offset.

% maybe mention something about how the vsini measurement doesn't take macroturbulence into account

\subsubsection{CHIRON Spectroscopy}\label{subsubsec:chiron}

We observed the remaining star, TOI-4773, using the CHIRON instrument: a fiber-fed, high-resolution \'echelle spectrograph installed on one of the 1.5-m Small and Moderate Aperture Research Telescope System (SMARTS) telescopes at the Cerro Tololo Inter-American Observatory (CTIO). We obtained the spectra of TOI-4773 using an image slicer with a resolving power of $\sim$ 80,000. Before and after collecting each science spectrum, we obtained Thorium-Argon calibration spectra. The RVs were then derived by using the least-squares deconvolution \citep{Donati:1997, Zhou:2021} of the observed spectra against non-rotating synthetic templates generated using the ATLAS9 models \citep{kurucz:1992}, before fitting the resulting line profile with a rotational broadening kernel, as prescribed by \cite{Gray:2005}. Finally, as we did with the TRES spectra, we obtained estimates of the effective temperature, metallicity, surface gravity, and projected radial velocity of TOI-4773 using SPC \citep{Buchhave:2012}.

Our CHIRON observations of TOI-4773 spanned 800 days, between UT 2022 December 21 and UT 2025 February 27. Prior to these observations, two reconnaissance spectra of TOI-4773 were collected using the TRES instrument on UT 2022 January 13 and UT 2022 February 9; however, these observations were not included in our global fit as the gamma and jitter associated with the TRES instrument would be poorly constrained with only two observations. There is no evidence of a long-term RV trend in our CHIRON data.

\subsubsection{NEID Spectroscopy}\label{subsubsec:neid}

We also used the NEID spectrograph \citep{Schwab:2016, Halverson:2016} to conduct RV follow-up observations of TOI-5350. NEID is a fiber-fed \citep{Kanodia:2018, Kanodia:2023} and ultra-stable \citep{Stefannson:2016, Robertson:2019} spectrograph mounted on the WIYN 3.5-meter telescope at Kitt Peak National Observatory in Arizona. The observations were carried out from 2023 October 4 to 2024 January 19, spanning 107 days and yielding 13 RV measurements. Data were taken in high-resolution mode (R~$\sim$~110{,}000) with an exposure time of 1300 seconds.

The NEID spectra were processed using version 1.4.0 of the NEID Data Reduction Pipeline (\texttt{NEID-DRP})\footnote{\url{https://neid.ipac.caltech.edu/docs/NEID-DRP/}}, which employs the cross-correlation function method to extract radial velocities. The resulting median RV uncertainty is 12 m s$^{-1}$, and the median signal-to-noise ratio at 550 nm is 24. Barycentric-corrected velocities, derived from re-weighted spectral orders (\texttt{CCFRVMOD}), were retrieved from the NExScI NEID Archive\footnote{\url{https://neid.ipac.caltech.edu/}.}. The full set of RVs is provided in the online journal.

\subsection{Speckle Imaging}\label{subsec:speckle}

The existence of nearby stars on the sky, whether they are bound to the target star or simply a foreground or background star, can disrupt or mimic some of the signals of hot Jupiters. Blended light from unresolved eclipsing binaries can lead to photometric signals resembling those of transiting planets. Composite spectra can lead to multiple sets of absorption and emission features, confounding RV measurements. These circumstances can lead to false positive detections as well as incorrect planetary and stellar parameters \citep[e.g.,][]{Ciardi:2015, Furlan:2017b, Furlan:2020}. As nearly half of FGK stars, the targets of this survey, have at least one stellar companion \citep{Matson:2018}, the detection of these companions is of paramount importance. To find or rule out these companions, we use speckle interferometry, a high-resolution imaging technique where thousands of images with short exposure times are collected of the target system \citep{Howell:2021}. An average power spectrum is then computed for each image and compared to the power spectrum of a single standard star to ascertain whether or not the target system is also single. 

In this article, we collected eight speckle observations of our five target systems using three separate instruments. These observations are organized by instrument used and are described in further detail in the following sections. A summary of these observations is shown in Table \ref{tab:hri}. Only one of the target systems, TOI-5261, was found to have a nearby star, which is unlikely to be bound and has minimal impact on our analysis.

% Exoplanet host stars may have spatially close companions (bound or line of sight) which can create a false- positive transit signal if it is, for example, an eclipsing binary (EB). Such ``Third-light” flux contamination from the close companion star will lead to an underestimated planetary radius if not accounted for in the transit model (Ciardi et al. 2015), can cause non-detections of small planets residing with the same exoplanetary system (Lester et al., 2021), and will result in incorrect parameters for both the planet and its host star (Furlan and Howell, 2017, 2020). Additional, close, bound companion stars, which exist in nearly one-half of FGK type stars (Matson et al., 2018) while less so for M class stars, provide crucial information toward our understanding of exoplanetary formation, dynamics and evolution (Howell et al., 2021). Thus, to search for close-in bound companions unresolved in TESS or other ground-based follow-up observations, we obtained high-resolution imaging speckle observations of TOI-6420.
%
\begin{table*}
    \centering
    \caption{Summary of Speckle Imaging Observations}
    \label{tab:hri}
    \begin{tabular}{l l l l l l c}
        \hline
        Target & Telescope & Instrument & Filter & Contrast & Obs. Date (UT) & Detection?\textsuperscript{\dag} \\
        \hline
        TOI-4138 & SAI (2.5~m) & Speckle Polarimeter & $I_c$ & $\Delta$5.3 mag at 1\arcsec & 2022 Mar 14 & No \\
        TOI-4773 & SOAR (4.1~m) & HRCam & $I_c$ & $\Delta$6.6 mag at 1\arcsec & 2022 Apr 15 & No \\
        TOI-5261 & SOAR (4.1~m) & HRCam & $I_c$ & $\Delta$5.9 mag at 1\arcsec & 2022 Feb 22 & Yes \\
        --- & SAI (2.5~m) & Speckle Polarimeter & $I_c$ & $\Delta$5.9 mag at 1\arcsec & 2023 Dec 2 & No \\
        TOI-5350 & SAI (2.5~m) & Speckle Polarimeter & $I_c$ & $\Delta$7.6 mag at 1\arcsec & 2022 Dec 22 & No \\
        --- & SOAR (4.1~m) & HRCam & $I_c$ & $\Delta$4.9 mag at 1\arcsec & 2023 Jan 5 & No \\
        TOI-6420 & Gemini (8~m) & Zorro & 562 nm & $\Delta$5.41 mag at 0.5\arcsec & 2024 Mar 14 & No \\
        --- & Gemini (8~m) & Zorro & 832 nm & $\Delta$6.31 mag at 0.5\arcsec & 2024 Mar 14 & No \\
        \hline
    \end{tabular}
    \vspace{2mm}

    \begin{minipage}{0.95\linewidth}
        \footnotesize
        \textbf{Note:} All images and contrast curves are available on ExoFOP.\\
        \textsuperscript{\dag} Detection refers to a positive detection of a star within the field of view of the AO or speckle instrument, subject to the maximum contrast possible with the instrument in question.
    \end{minipage}
\end{table*}
\subsubsection{Sternberg Astronomical Institute Speckle Polarimeter}

TOI-5261 and TOI-5350 were observed on 2 December 2023 and 22 December 2022, respectively, with the speckle polarimeter on the 2.5-m telescope at the Caucasian Observatory of Sternberg Astronomical Institute (SAI) of Lomonosov Moscow State University. A low--noise CMOS detector Hamamatsu ORCA--quest \citep{Strakhov:2023} was used as a detector. \confirmedtoi was observed on 14 March 2022 with a previous, EMCCD--based, version of the instrument. The atmospheric dispersion compensator was active, which allowed using the $I_\mathrm{c}$ band. The respective angular resolution is $0.083\arcsec$. No companions were detected for either of these targets. The detection limits at distances $0.25$ and $1.0^{\prime\prime}$ from the star are for \confirmedtoi\ - $\Delta I_\mathrm{c}=3.9$ mag and $5.3$ mag, for TOI-5261 - $\Delta I_\mathrm{c}=3.0$ mag and $5.8$ mag, and for TOI-5350 - $\Delta I_\mathrm{c}=3.3$ mag and $7.6$ mag.

\subsubsection{Southern Astrophysical Research Telescope HRCam Imager}

% High-angular resolution imaging is needed to search for nearby sources that can contaminate the TESS photometry, resulting in an underestimated planetary radius, or be the source of astrophysical false positives, such as background eclipsing binaries.

 We searched for stellar companions to TOIs 4773, 5261, and 5350 with speckle imaging on the 4.1-m Southern Astrophysical Research (SOAR) telescope \citep{Tokovinin:2018} on 15 April and 22 February 2022 UT and 5 January 2023 UT, respectively, observing in Cousins I-band, a similar visible bandpass as TESS. More details of the observations within the SOAR TESS survey are available in \cite{Ziegler:2020}. The 5$\sigma$ detection sensitivity and speckle auto-correlation functions from the observations are available on ExoFOP\footnote{\url{https://exofop.ipac.caltech.edu/tess/}}. No nearby stars were detected within 3\arcsec of TOI-4473 and TOI-5350 in the SOAR observations. A faint star, 5.9 magnitudes dimmer than TOI-5261, was detected at approximately 1.8\arcsec separation from TOI-5261. This nearby star appears in the Gaia DR3 catalog and appears not to be comoving with the target star based on the Gaia proper motion estimates, and thus is likely to be an unbound asterism.

\subsubsection{Gemini-South Zorro Imager}

TOI-6420 was observed on 2024 March 14 UT using the Zorro speckle instrument on the Gemini South 8-m telescope\footnote{\url{https://www.gemini.edu/sciops/instruments/alopeke-zorro/}} \citep{Scott:2021}.  Zorro provides simultaneous speckle imaging in two bands (562nm and 832 nm) with output data products including a reconstructed image with robust contrast limits on companion detections. Nine sets of 1000 X 0.06 sec exposures were collected and subjected to Fourier analysis in our standard reduction pipeline (see \citealt{Howell:2011}). We find that TOI-6420 is a single star with no companion brighter than 5-7 magnitudes below that of the target star from the diffraction limit (20 mas) out to 1.2”. At the distance of TOI-6420 (d=635 pc) these angular limits correspond to spatial limits of  13 to  762 au.

\subsection{Archival Photometry}

In order to expand our wavelength coverage of the target stars and constrain their spectral energy distributions (SEDs), we queried VizieR\footnote{\url{https://vizier.cfa.harvard.edu/viz-bin/VizieR-2}} \citep{Ochsenbein:2000} to obtain archival photometry from Gaia DR3 \citep{GaiaDR3}, the Two Micron All-Sky Survey (2MASS; \citealt{Cutri:2003, Skrutskie:2006}), and the Wide-field Infrared Survey Explorer (WISE; \citealt{Wright:2010, Cutri:2012}). The retrieved bandpasses included Gaia $G$, $G_{\rm BP}$, $G_{\rm RP}$, 2MASS $J$, $H$, $K_{\rm s}$, and WISE $W1$, $W2$, and $W3$, which span a wavelength range of $ 0.33 - 17 \mu$m. To ensure that our uncertainties are not underestimated, we adopt a systematic floor on the uncertainty of these magnitudes as described in \cite{Eastman:2019}. These values and adopted uncertainties are reported in Table \ref{tab:lit}, along with astrometric parameters selected from Gaia DR3. 

\providecommand{\bjdtdb}{\ensuremath{\rm {BJD_{TDB}}}}
\providecommand{\feh}{\ensuremath{\left[{\rm Fe}/{\rm H}\right]}}
\providecommand{\teff}{\ensuremath{T_{\rm eff}}}
\providecommand{\teq}{\ensuremath{T_{\rm eq}}}
\providecommand{\ecosw}{\ensuremath{e\cos{\omega_*}}}
\providecommand{\esinw}{\ensuremath{e\sin{\omega_*}}}
\providecommand\msun{M$_\odot$\xspace}
\providecommand{\rsun}{R$_\odot$\xspace}
\providecommand{\lsun}{L$_\odot$\xspace}
\providecommand{\mj}{\ensuremath{\,M_{\rm J}}}
\providecommand{\rj}{\ensuremath{\,R_{\rm J}}}
\providecommand{\me}{\ensuremath{\,M_{\rm E}}}
\providecommand{\re}{\ensuremath{\,R_{\rm E}}}
\providecommand{\fave}{\langle F \rangle}
\providecommand{\fluxcgs}{10$^9$ erg s$^{-1}$ cm$^{-2}$}
\providecommand{\tess}{\textit{TESS}\xspace}
\begin{table*}
    \centering
    \caption{Measured Properties from Literature}
    \label{tab:lit}
    \resizebox{\textwidth}{!}{
    \begin{tabular}{l l c c c c c c}
        \hline
         & & TOI-4138 & TOI-4773 & TOI-5261 & TOI-5350 & TOI-6420 & Source \\
        \hline
        \multicolumn{8}{l}{\textbf{Other identifiers}:} \\
        & TESS Input Catalog & TIC 257060897 & TIC 415276070 & TIC 402828941 & TIC 68808155 & TIC 143526233 & \\
        & TYCHO-2 & TYC 4417-1588-1 & TYC 5992-2066-1 & --- & --- & --- & \\
        & 2MASS & J15100767+7242372 & J07394624-2129081 & J20215006+1926094 & J04573656+2136074 & J08221064-1907274 & \\
        & Gaia DR3 & 1697129530714536320 & 5715145275211633792 & 1816307623283205248 & 3412163401401508096 & 5707456527828738048 & \\
        \hline
        \multicolumn{8}{l}{\textbf{Astrometric Parameters}:} \\
        $\alpha_{J2000}\ddagger$ & Right Ascension (h:m:s) & 15:10:07.67 & 07:39:46.244 & 20:21:50.179 & 04:57:36.554 & 08:22:10.643  & 1 \\
        $\delta_{J2000}\ddagger$ & Declination (d:m:s) & 72:42:37.247 & -21:29:08.201 & 19:26:09.099 & 21:36:07.487 & -19:07:27.456  & 1 \\
        $\mu_{\alpha}$ & Gaia DR3 proper motion in RA (mas yr$^{-1}$)& $13.508 \pm 0.013$ & $-7.863 \pm 0.014$ & $-4.202 \pm 0.248$ & $-1.591 \pm 0.037$ & $-8.852 \pm 0.013$  & 1 \\
        $\mu_{\delta}$ & Gaia DR3 proper motion in Dec (mas yr$^{-1}$)& $-7.779 \pm 0.015$ & $-1.613 \pm 0.017$ & $-3.497 \pm 0.386$ & $-24.112 \pm 0.024$ & $1.893 \pm 0.014$  & 1 \\
        $\pi$ & Gaia DR3 Parallax (mas) & $1.9695 \pm 0.0112$ & $1.9493 \pm 0.0163$ & $-0.2016 \pm 0.3236$ & $3.5179 \pm 0.0289$ & $1.6413 \pm 0.0166$  & 1 \\
        $v\sin{i_\star}$ & Projected rotational velocity (km s$^{-1}$) & $5.45 \pm 0.074$ & $15.3 \pm 0.29$ & $5.73 \pm 0.11$ & $9.01 \pm 0.13$ & $6.42 \pm 0.088$  & 2 \\
        \multicolumn{8}{l}{\textbf{Photometric Parameters}:} \\
        ${\rm G}$ & Gaia $G$ mag. & $11.662 \pm 0.020$ & $11.714 \pm 0.020$ & $12.602 \pm 0.020$ & $11.702 \pm 0.020$ & $12.890 \pm 0.020$  & 1 \\
        $G_{\rm BP}$ & Gaia $G_{\rm BP}$ mag. & $11.959 \pm 0.020$ & $11.936 \pm 0.020$ & $12.968 \pm 0.020$ & $12.108 \pm 0.020$ & $13.199 \pm 0.020$  & 1 \\
        $G_{\rm RP}$ & Gaia $G_{\rm RP}$ mag. & $11.202 \pm 0.020$ & $11.340 \pm 0.020$ & $12.093 \pm 0.020$ & $11.117 \pm 0.020$ & $12.422 \pm 0.020$  & 1 \\
        ${\rm T}$ & TESS mag. & $11.2633 \pm 0.007$ & $11.4048 \pm 0.006$ & $12.1444 \pm 0.006$ & $11.1846 \pm 0.006$ & $12.4877 \pm 0.006$  & 3 \\
        $J$ & 2MASS $J$ mag. & $10.696 \pm 0.021$ & $10.941 \pm 0.022$ & $11.525 \pm 0.024$ & $10.412 \pm 0.022$ & $11.893 \pm 0.024$  & 4 \\
        $H$ & 2MASS $H$ mag. & $10.446 \pm 0.020$ & $10.765 \pm 0.026$ & $11.234 \pm 0.030$ & $10.072 \pm 0.020$ & $11.638 \pm 0.025$  & 4 \\
        $K$ & 2MASS $K$ mag. & $10.387 \pm 0.020$ & $10.715 \pm 0.021$ & $11.156 \pm 0.020$ & $9.985 \pm 0.020$ & $11.615 \pm 0.024$  & 4 \\
        $W1$ & WISE $W1$ mag. & $10.348 \pm 0.030$ & $10.665 \pm 0.030$ & $11.110 \pm 0.030$ & $9.925 \pm 0.030$ & $11.532 \pm 0.030$  & 5 \\
        $W2$ & WISE $W2$ mag. & $10.375 \pm 0.030$ & $10.696 \pm 0.030$ & $11.168 \pm 0.030$ & $9.935 \pm 0.030$ & $11.557 \pm 0.030$  & 5 \\
        $W3$ & WISE $W3$ mag. & $10.399 \pm 0.053$ & $10.793 \pm 0.088$ & $11.362 \pm 0.222$ & $9.928 \pm 0.074$ & $11.452 \pm 0.177$  & 5 \\
        \hline
    \end{tabular}
    } % end resizebox
    \vspace{2mm}
    \begin{minipage}{0.95\linewidth}
        \footnotesize
        \textbf{Notes:} The uncertainties of the photometric measurements have a systematic floor applied that is usually larger than the reported catalog errors.\\
        $\ddagger$ Right Ascension and Declination are in epoch J2000. Coordinates are from Vizier where Gaia RA and Dec have been precessed and corrected from epoch J2016.\\
        Sources: (1) \cite{GaiaDR3}; (2) \S\ref{subsubsec:tres} \& \S\ref{subsubsec:chiron}; (3) \cite{Stassun:2019}; (4) \cite{Cutri:2003, Skrutskie:2006}; (5) \cite{Wright:2010, Cutri:2012}
    \end{minipage}
\end{table*}

\section{Analysis}\label{sec:analysis}

In addition to confirming that each planet is real, we aim to carefully characterize each planetary system in this article. In the case of \confirmedtoi, our aim is to reanalyze this benchmark system using the methodologies of the MEEP survey to ensure self-consistency in the final sample. Therefore, we treat it as if it were a new planet discovery and fit it in the same fashion as the other systems in this work, without using the results of its discovery paper \citep{Montalto:2022} as priors. After characterizing each planetary system, we consider the additional information not included in our fits to further investigate each system.

\subsection{\exofast Global Fits}\label{subsec:exofast}

Following the procedure used in the first paper in this series \citep{Schulte:2024}, we used the open-source global fitting software \exofast\footnote{\url{https://github.com/jdeast/EXOFASTv2}} \citep{Eastman:2013, Eastman:2019} to estimate the properties of the stars and planets in this work. \exofast is a differential evolution Markov Chain Monte Carlo (MCMC) code written in \texttt{IDL}. Uniquely, it has the capability to fit the stellar spectral energy distribution (SED) using MESA Isochrones and Stellar Tracks \citep[MIST;][]{Paxton:2011}, radial velocities (RVs), and planetary transits simultaneously to ensure a self-consistent set of planetary and stellar parameters. To ensure that our MCMC chains were well-mixed, that the parameter space was properly explored, and that the best solution was found, we adopted a strict set of convergence criteria: the Gelman-Rubin statistic \citep{Ford:2006} must be smaller than 1.01 and the number of independent draws must be larger than 1000.

In order to properly account for past observations of these systems, we applied several Bayesian priors to our global fits. We adopted Gaussian priors on the parallax of each target from Gaia Data Release 3 \citep{GaiaDR3}, which were corrected according to \cite{Lindegren:2021}. Additionally, we placed Gaussian priors on the metallicity of each star, with the mean of the Gaussian set to a weighted average of the spectroscopic metallicities from TRES and CHIRON (as described in \S \ref{subsubsec:tres} and \S \ref{subsubsec:chiron}), where the signal-to-noise resolution ratio (SNR) per resolution element was used as the weight. The standard deviation of the spectroscopic metallicities was used as the width of the Gaussian. Finally, in the initial fit that was run for each system, we accounted for possible incomplete correction of the light contamination in the \tess target pixel by fitting for a dilution term. We then placed a Gaussian prior, centered at 0\%, with a standard deviation that is 10\% of the dilution factor, $D$, defined as $D = C/(1 + C)$, where $C$ is the contamination ratio from the \tess Input Catalog (TIC) v8.2 \citep{Stassun:2018, Stassun:2019}. This prior is meant to represent the general assumption that the QLP and SPOC pipelines have correctly accounted for the light contamination in the target pixel. After the first global fit was run to convergence, if the follow-up lightcurves were in good agreement with the \tess lightcurves and the median of the dilution term was consistent within $1\sigma$ with zero, we removed the dilution term from the fit to avoid inflating our uncertainties. The only system in which we left the dilution term in was TOI-4773, which had a \tess dilution term with a median value of $-0.037 \pm 0.025$. Finally, we placed a conservative upper limit on the V-band dust extinction of each target using the \cite{Schlafly:2011} dust maps in order to constrain the stellar radius. These priors and constraints are listed at the top of Table \ref{tab:median}. In addition to these priors, we adopted starting values on the stellar mass, stellar radius, and stellar effective temperature from the TIC. The starting values for the transit epoch $T_C$, orbital period $P$, and the ratio of planetary and stellar radii $R_P/R_*$ were retrieved from the \tess mission catalog on ExoFOP\footnote{\url{https://exofop.ipac.caltech.edu/tess/}}.

\providecommand{\bjdtdb}{\ensuremath{\rm {BJD_{TDB}}}}
\providecommand{\feh}{\ensuremath{\left[{\rm Fe}/{\rm H}\right]}}
\providecommand{\teff}{\ensuremath{T_{\rm eff}}}
\providecommand{\teq}{\ensuremath{T_{\rm eq}}}
\providecommand{\ecosw}{\ensuremath{e\cos{\omega_*}}}
\providecommand{\esinw}{\ensuremath{e\sin{\omega_*}}}
\providecommand{\msun}{\ensuremath{\,M_\Sun}}
\providecommand{\rsun}{\ensuremath{\,R_\Sun}}
\providecommand{\lsun}{\ensuremath{\,L_\Sun}}
\providecommand{\mj}{\ensuremath{\,M_{\rm J}}}
\providecommand{\rj}{\ensuremath{\,R_{\rm J}}}
\providecommand{\me}{\ensuremath{\,M_{\rm E}}}
\providecommand{\re}{\ensuremath{\,R_{\rm E}}}
\providecommand{\fave}{\langle F \rangle}
\providecommand{\fluxcgs}{10$^9$ erg s$^{-1}$ cm$^{-2}$}
\providecommand{\tess}{\textit{TESS}\xspace}
\begin{table*}
\centering
\caption{Median Values and 68\% Confidence Intervals for Fitted Stellar and Planetary Parameters}
\label{tab:median}
\scriptsize
\begin{tabular}{lllcccccc}
\hline
&  & TOI-4138* & TOI-4773 & TOI-5261 & TOI-5350 & TOI-6420\\
\hline
\multicolumn{7}{l}{\textbf{Priors}:} \\
$\pi$ & Gaia Parallax (mas)& $\mathcal{G}$[2.0145, 0.01501] & $\mathcal{G}$[1.9904, 0.0316] & $\mathcal{G}$[2.655, 0.06138] & $\mathcal{G}$[3.5598, 0.03058] & $\mathcal{G}$[1.666, 0.01938] \\
$[{\rm Fe/H}]$ & Metallicity (dex)& $\mathcal{G}$[0.062267, 0.093633] & $\mathcal{G}$[0.1471, 0.165] & $\mathcal{G}$[0.25328, 0.078102] & $\mathcal{G}$[0.021428, 0.10076] & $\mathcal{G}$[0.25109, 0.077843] \\
$A_V$ & V-band extinction (mag)& $\mathcal{U}$[0, 0.089406] & $\mathcal{U}$[0, 2.9664] & $\mathcal{U}$[0, 0.66265] & $\mathcal{U}$[0, 1.1733] & $\mathcal{U}$[0, 0.2852] \\
$D_T$ & Dilution in \tess& --- & $\mathcal{G}$[0, 0.030469] & --- & --- & --- \\
\hline
\multicolumn{7}{l}{\textbf{Stellar Parameters}:} \\
$M_*$ & Mass (\msun) & $1.187^{+0.110}_{-0.058}$ & $1.452^{+0.082}_{-0.075}$ & $1.037^{+0.053}_{-0.057}$ & $1.167^{+0.063}_{-0.066}$ & $1.165^{+0.069}_{-0.078}$ \\
$R_*$ & Radius (\rsun) & $1.863^{+0.050}_{-0.054}$ & $1.55 \pm 0.046$ & $1.021^{+0.033}_{-0.032}$ & $1.263^{+0.037}_{-0.030}$ & $1.312^{+0.051}_{-0.048}$ \\
$L_*$ & Luminosity (\lsun) & $4.03 \pm 0.15$ & $4.45^{+0.72}_{-0.52}$ & $1.049^{+0.120}_{-0.091}$ & $2.16^{+0.17}_{-0.14}$ & $2.05^{+0.14}_{-0.16}$ \\
$\rho_*$ & Density (cgs) & $0.259^{+0.033}_{-0.022}$ & $0.55^{+0.051}_{-0.045}$ & $1.37^{+0.14}_{-0.13}$ & $0.821^{+0.062}_{-0.080}$ & $0.726^{+0.100}_{-0.097}$ \\
$\log{g}$ & Surface gravity (cgs) & $3.972^{+0.044}_{-0.029}$ & $4.22^{+0.029}_{-0.028}$ & $4.436^{+0.031}_{-0.033}$ & $4.304^{+0.025}_{-0.035}$ & $4.268^{+0.044}_{-0.048}$ \\
$T_{\rm eff}$ & Effective temperature (K) & $5993^{+89}_{-83}$ & $6740^{+250}_{-200}$ & $5790^{+140}_{-120}$ & $6220^{+130}_{-120}$ & $6020^{+140}_{-150}$ \\
$[{\rm Fe/H}]$ & Metallicity (dex) & $0.054^{+0.078}_{-0.070}$ & $0.11^{+0.14}_{-0.12}$ & $0.254^{+0.076}_{-0.077}$ & $0.044^{+0.072}_{-0.059}$ & $0.24^{+0.074}_{-0.077}$ \\
$[{\rm Fe/H}]_{0}$ & Initial metallicity & $0.098^{+0.072}_{-0.068}$ & $0.238^{+0.100}_{-0.090}$ & $0.24^{+0.073}_{-0.074}$ & $0.1^{+0.062}_{-0.057}$ & $0.26 \pm 0.066$ \\
Age & Age (Gyr) & $5.6^{+1.2}_{-1.7}$ & $1.07^{+0.75}_{-0.55}$ & $4.2^{+3.4}_{-2.3}$ & $3.1^{+1.9}_{-1.4}$ & $4.4^{+2.8}_{-20}$ \\
EEP & Equivalent evolutionary phase & $450.9^{+5.8}_{-360}$ & $334^{+13}_{-20}$ & $351^{+37}_{-21}$ & $367^{+36}_{-27}$ & $395^{+31}_{-46}$ \\
$A_V$ & V-band extinction (mag) & $0.061^{+0.021}_{-0.034}$ & $0.21^{+0.15}_{-0.13}$ & $0.16^{+0.12}_{-0.10}$ & $0.761^{+0.078}_{-0.075}$ & $0.183^{+0.073}_{-0.100}$ \\
$d$ & Distance (pc) & $496.9 \pm 3.7$ & $502.8^{+80}_{-7.7}$ & $376.7^{+8.6}_{-8.3}$ & $280.7 \pm 2.4$ & $600.5^{+70}_{-6.9}$ \\
\multicolumn{7}{l}{\textbf{Planetary Parameters}:} \\
$P$ & Period (days) & $3.66003639^{+0.00000099}_{-0.00000098}$ & $1.7452851 \pm 0.0000014$ & $4.1509768 \pm 0.0000068$ & $7.581367^{+0.000012}_{-0.000011}$ & $6.961508 \pm 0.000012$ \\
$R_{\rm P}$ & Radius (\rj) & $1.54^{+0.043}_{-0.045}$ & $1.358^{+0.049}_{-0.048}$ & $1.035^{+0.042}_{-0.041}$ & $1.108^{+0.038}_{-0.029}$ & $1.065^{+0.065}_{-0.059}$ \\
$M_{\rm P}$ & Mass (\mj) & $0.634^{+0.041}_{-0.033}$ & $5.31^{+0.35}_{-0.34}$ & $11.49^{+0.43}_{-0.46}$ & $6.59^{+0.25}_{-0.26}$ & $8.2^{+0.36}_{-0.40}$ \\
$T_C$ & Time of conjunction (\bjdtdb) & $2459660.60712 \pm 0.00017$ & $2459251.18071^{+0.00049}_{-0.00050}$ & $2459442.96658 \pm 0.00058$ & $2459520.68127 \pm 0.00040$ & $2459985.8467 \pm 0.0012$ \\
$T_0$ & Optimal conjunction time (\bjdtdb) & $2459473.94542 \pm 0.00013$ & $2459673.53995 \pm 0.00028$ & $2459716.931 \pm 0.00036$ & $2459611.65813 \pm 0.00038$ & $2459909.2703 \pm 0.0012$ \\
$a$ & Semi-major axis (AU) & $0.04922^{+0.00140}_{-0.00081}$ & $0.03216^{+0.00060}_{-0.00056}$ & $0.05134^{+0.00086}_{-0.00096}$ & $0.0796^{+0.0014}_{-0.0015}$ & $0.0753^{+0.0014}_{-0.0017}$ \\
$i$ & Inclination (Degrees) & $84.55^{+0.44}_{-0.40}$ & $78.64^{+0.61}_{-0.56}$ & $86.07^{+0.29}_{-0.30}$ & $89.04^{+0.59}_{-0.54}$ & $85.8^{+0.27}_{-0.30}$ \\
$e$ & Eccentricity & $0.039^{+0.038}_{-0.023}$ & $0.024^{+0.027}_{-0.017}$ & $0.1585^{+0.0100}_{-0.0078}$ & $0.0165^{+0.0120}_{-0.0081}$ & $0.078^{+0.014}_{-0.013}$ \\
$\omega_*$ & Argument of periastron (Degrees) & $-48^{+60}_{-24}$ & $-113^{+87}_{-100}$ & $32.4^{+5.4}_{-4.8}$ & $121^{+42}_{-21}$ & $48.4^{+9.4}_{-110}$ \\
$\teq$ & Equilibrium temperature (K) & $1775^{+22}_{-25}$ & $2255^{+73}_{-59}$ & $1244^{+30}_{-25}$ & $1195^{+21}_{-19}$ & $1213^{+20}_{-21}$ \\
$\tau_{\rm circ}$ & Tidal circularization timescale (Gyr) & $0.0675^{+0.0100}_{-0.0088}$ & $0.0498^{+0.0097}_{-0.0083}$ & $10.8^{+2.5}_{-2.1}$ & $87^{+12}_{-15}$ & $84^{+31}_{-24}$ \\
$K$ & RV semi-amplitude (m/s) & $74.1 \pm 2.9$ & $684 \pm 37$ & $1424^{+19}_{-21}$ & $613.3^{+7.8}_{-6.9}$ & $785^{+15}_{-16}$ \\
$R_{\rm P}/R_*$ & Radius of planet in stellar radii  & $0.08499^{+0.00034}_{-0.00035}$ & $0.0901 \pm 0.0014$ & $0.1041 \pm 0.0013$ & $0.09027^{+0.00072}_{-0.00068}$ & $0.0835^{+0.0027}_{-0.0025}$ \\
$a/R_*$ & Semi-major axis in stellar radii  & $5.68^{+0.23}_{-0.16}$ & $4.46^{+0.14}_{-0.13}$ & $10.81^{+0.36}_{-0.35}$ & $13.58^{+0.33}_{-0.46}$ & $12.33^{+0.56}_{-0.57}$ \\
Depth & \tess flux decrement at mid-transit & $0.00785^{+0.000045}_{-0.000044}$ & $0.00751^{+0.00023}_{-0.00022}$ & $0.01153 \pm 0.00023$ & $0.00911 \pm 0.00014$ & $0.00649^{+0.00033}_{-0.00032}$ \\
$\tau$ & Ingress/egress transit duration (days) & $0.02187^{+0.00090}_{-0.00089}$ & $0.0275^{+0.0024}_{-0.0020}$ & $0.0157^{+0.0013}_{-0.0012}$ & $0.01624^{+0.00130}_{-0.00074}$ & $0.0279^{+0.0043}_{-0.0034}$ \\
$T_{14}$ & Total transit duration (days) & $0.19757^{+0.00081}_{-0.00080}$ & $0.0817 \pm 0.0012$ & $0.0983^{+0.0014}_{-0.0013}$ & $0.1869^{+0.0015}_{-0.0012}$ & $0.1146^{+0.0035}_{-0.0034}$ \\
$b$ & Transit impact parameter & $0.556^{+0.022}_{-0.025}$ & $0.8857^{+0.0066}_{-0.0072}$ & $0.666^{+0.028}_{-0.032}$ & $0.22^{+0.12}_{-0.14}$ & $0.849^{+0.018}_{-0.020}$ \\
$\rho_{\rm P}$ & Density (cgs) & $0.215^{+0.024}_{-0.019}$ & $2.63^{+0.33}_{-0.29}$ & $12.9^{+1.7}_{-1.5}$ & $6.03^{+0.49}_{-0.63}$ & $8.4^{+1.7}_{-1.5}$ \\
$\log{g_{\rm P}}$ & Surface gravity (cgs) & $2.821^{+0.037}_{-0.032}$ & $3.854 \pm 0.039$ & $4.425^{+0.036}_{-0.037}$ & $4.125^{+0.024}_{-0.035}$ & $4.253^{+0.056}_{-0.060}$ \\
$M_{\rm P}/M_*$ & Mass ratio  & $0.000507^{+0.000023}_{-0.000024}$ & $0.00349 \pm 0.00020$ & $0.01059^{+0.00025}_{-0.00024}$ & $0.0054^{+0.00013}_{-0.00012}$ & $0.00673^{+0.00021}_{-0.00019}$ \\
$d/R_*$ & Separation at mid-transit  & $5.8^{+0.49}_{-0.30}$ & $4.49^{+0.25}_{-0.19}$ & $9.71^{+0.38}_{-0.40}$ & $13.38^{+0.43}_{-0.50}$ & $11.58^{+0.59}_{-0.60}$ \\
\hline
\end{tabular}
\begin{flushleft}
\textbf{Notes:} The priors for each system are labeled as $\mathcal{G}$[mean, standard deviation] if they are Gaussian priors and $\mathcal{U}$[lower limit, upper limit] if they are uniform priors. *TOI-4138's posterior is bimodal in stellar mass and age. This table presents the median values; see Table~\ref{tab:bimodal} for the split solutions.
\end{flushleft}
\end{table*}

In the initial fits for each system, we included a linear slope term in the radial velocity fit to account for additional companions that do not transit. After the first fit is run to convergence, if this slope term is consistent within $1\sigma$ with zero, we remove the term for our final fit to ensure that the minimum number of free parameters is used. The systems with nonzero RV slopes are possible hosts to additional companions and are compelling targets for future RV monitoring. With the exception of \confirmedtoi (see \S \ref{subsubsec:tres}), only TOI-5261 had tentative evidence of a linear trend in its RVs, with a median linear slope of $-0.050^{+0.043}_{-0.047}$~m s$^{-1}$ day$^{-1}$. This is only barely inconsistent within 1 $\sigma$ with zero, and there is no evidence of a turnover in the RVs that would allow us to fit a second planet. Therefore, this system is merely a more compelling target for future follow-up to determine if a longer baseline can help to constrain the parameters of an additional, more distant companion.

Finally, in all of our \exofast fits, eccentricity was allowed to float as a free parameter. Eccentricity is primarily determined from the RV fit, but is difficult to precisely constrain when a planet's orbit is nearly circular. Although eccentricity, $e$, and the argument of periastron, $\omega_\star$, are parameterized as $\sqrt{e}\cos{\omega_\star}$ and $\sqrt{e}\sin{\omega_\star}$, this parameterization does not eliminate the Lucy-Sweeney bias \citep{Lucy:1971}. The Lucy-Sweeney bias states that a parameter, such as eccentricity, which is only allowed to be positive is often biased towards larger values than the true value, when that true value is close to zero. As this is also dictated by the certainty of the fit, \cite{Eastman:2019} states that the eccentricity of a planet must be 2.45$\sigma$ greater than zero in order to be 95\% confident that the orbit is eccentric. The planets in this sample, owing to their large masses and the precisions of the instruments that obtained their RVs, have reasonably precise eccentricities. However, due to the Lucy-Sweeney bias, only TOI-5261 b and TOI-6420 b have significant eccentricity. The other three planets have measured eccentricities that are consistent with a circular orbit. These eccentricities, along with the medians and uncertainties for other fitted stellar and planetary parameters for each system, are presented in Table \ref{tab:median}.

\subsection{Lithium Equivalent Width Measurements}\label{subsec:lithium}

Lithium is destroyed by proton- and $\alpha$-capture reactions in stellar interiors where temperatures exceed $\sim 3 \times 10^6$ K \cite{Bodenheimer:1965}. In low-mass stars where convection plays a major role in the transport of material to the interior of the star, Li can be transported from the photosphere to the interior where it is destroyed, leading to the depletion of Li in spectra of the star's surface. Therefore, the abundance of Li has been used as a tracer for the ages of young stars with temperatures $< 6500$ K \cite[e.g.,][]{Jeffries:2023}.

\begin{figure*}
    \centering
    \includegraphics[width=0.45\textwidth,height=\textheight,keepaspectratio]{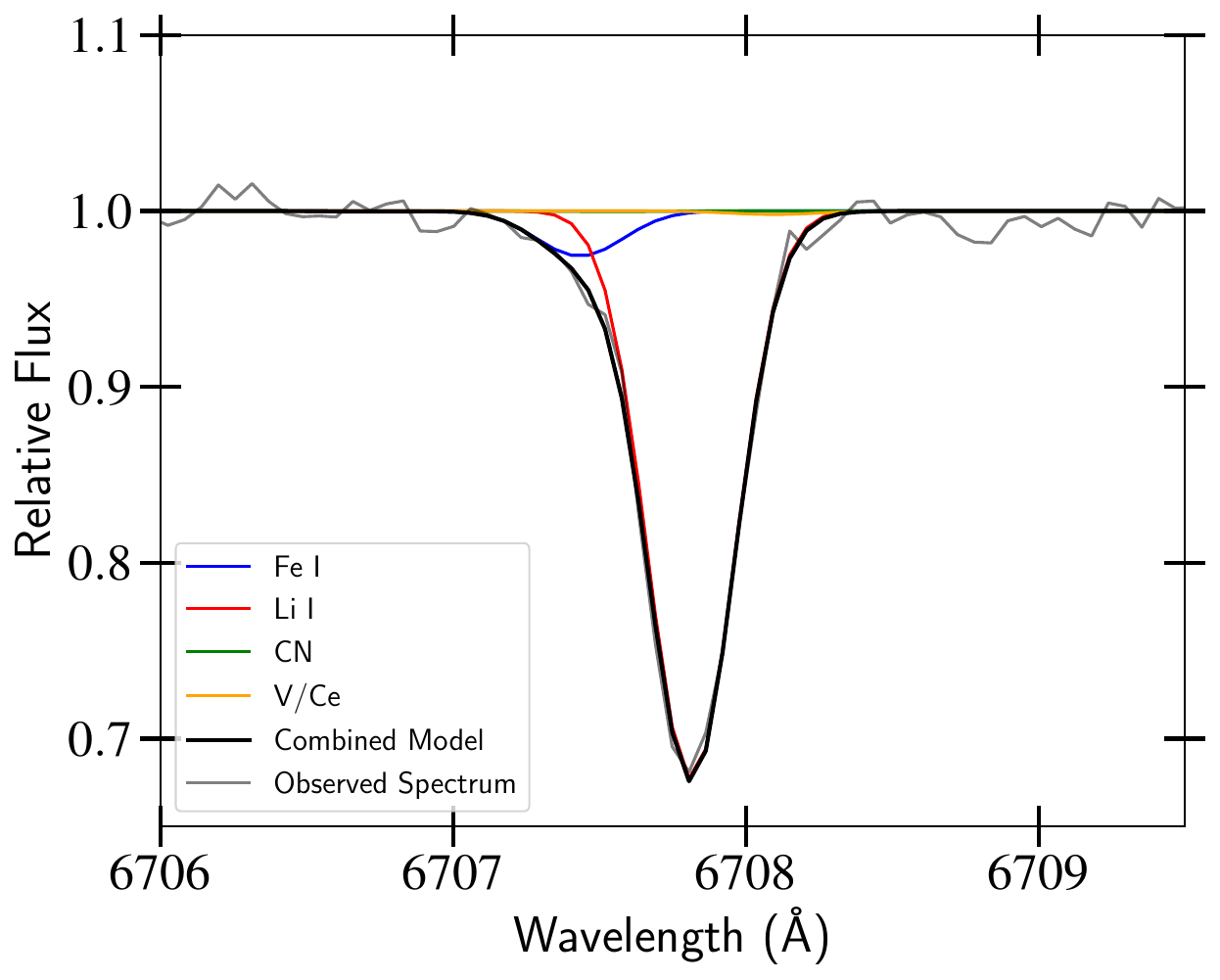}
    \caption{32 co-added TRES spectra of \confirmedtoi, illustrating the lithium doublet at 6707.814 \AA. Five Gaussians were fit to account for possible blending features of Fe, CN, V, and Ce in addition to the Li feature. The only features that contribute significantly to \confirmedtoi's spectrum are Fe I and Li I. The equivalent width of the Li I feature was measured to be $120. \pm 13$ m\AA.}
    \label{fig:li_ew_toi4138}
\end{figure*}

Four of the five host stars in this sample (\confirmedtoi, TOI-4773, TOI-5350, and TOI-6420) were found to have detectable lithium absorption features in their spectra. To characterize the strength of their Li features and the quantity of Li in their photospheres, we measured the equivalent width (EW) of the Li I absorption doublet at 6707.8 \AA\ in each star's spectrum and then compared the Li EW to a control sample of stars from the GALAH survey \citep{Buder:2024}. First, we used the \texttt{specutils}\footnote{\url{https://github.com/astropy/specutils}} Python package to remove the blaze function and RV shift from each observed spectrum and then co-added them to increase the signal-to-noise ratio. Then, we followed methodologies similar to those used in \cite{Wang:2024} to measure the Li EW. We construct five Gaussians to the absorption feature to account for the Li I feature (6707.814 \AA) along with several blending features: CN (6706.730 \AA\ and 6707.545 \AA), Fe I (6707.433 \AA), and two features of V and Ce that are near enough in wavelength and strength that we model them as one Gaussian feature at 6708.096 \AA. We hold the widths of all lines to be constant, as the line shape is dominated by instrumental and rotational broadening. Then, we perform a least squares fit using the \texttt{scipy.optimize.curve\_fit} function \citep{Virtanen:2020}. The results of our fits for each star are presented in Table \ref{tab:lithium} and an example of the EW calculation for the most prominent Li feature in \confirmedtoi is shown in Figure \ref{fig:li_ew_toi4138}.

\begin{table*}
\centering
\caption{Lithium measurements of host stars.}
\label{tab:lithium}
\scriptsize
\begin{tabular}{l c c c c}
\hline
 & TOI-4138 & TOI-4773 & TOI-5350 & TOI-6420 \\
\hline
Li EW [m\AA] & $120. \pm 13$ & $43.1 \pm 11.6$ & $41.1 \pm 9.5$ & $45.8 \pm 21.5$ \\
GALAH DR4 Baseline Li EW [m\AA] & $33.0 \pm 13.1$ & $11.9 \pm 7.6$ & $36.0 \pm 8.2$ & $34.6 \pm 9.3$ \\
GALAH DR4 Control Sample Size & 1381 & 177 & 3848 & 3883 \\
Li EW Modified Z-score & 4.47 & 2.79 & 0.419 & 0.809 \\
Li EW Percentile Rank & 99.86\% & 80.23\% & 68.30\% & 82.67\% \\
\texttt{PyMOOGi} A(Li) [dex] & 2.63 & 2.78 & 2.31 & 2.24 \\
\texttt{stardis} A(Li) [dex] & $2.54 \pm 0.06$ & $2.08 \pm 0.04$ & $1.83 \pm 0.05$ & $1.78 \pm 0.06$ \\
\hline
\end{tabular}
\begin{flushleft}
\textbf{Note:} The GALAH DR4 baseline Li EW values are reported as [median] $\pm$ [median absolute deviation].
\end{flushleft}
\end{table*}

\begin{figure*}
    \centering
    \includegraphics[width=0.45\textwidth,height=\textheight,keepaspectratio]{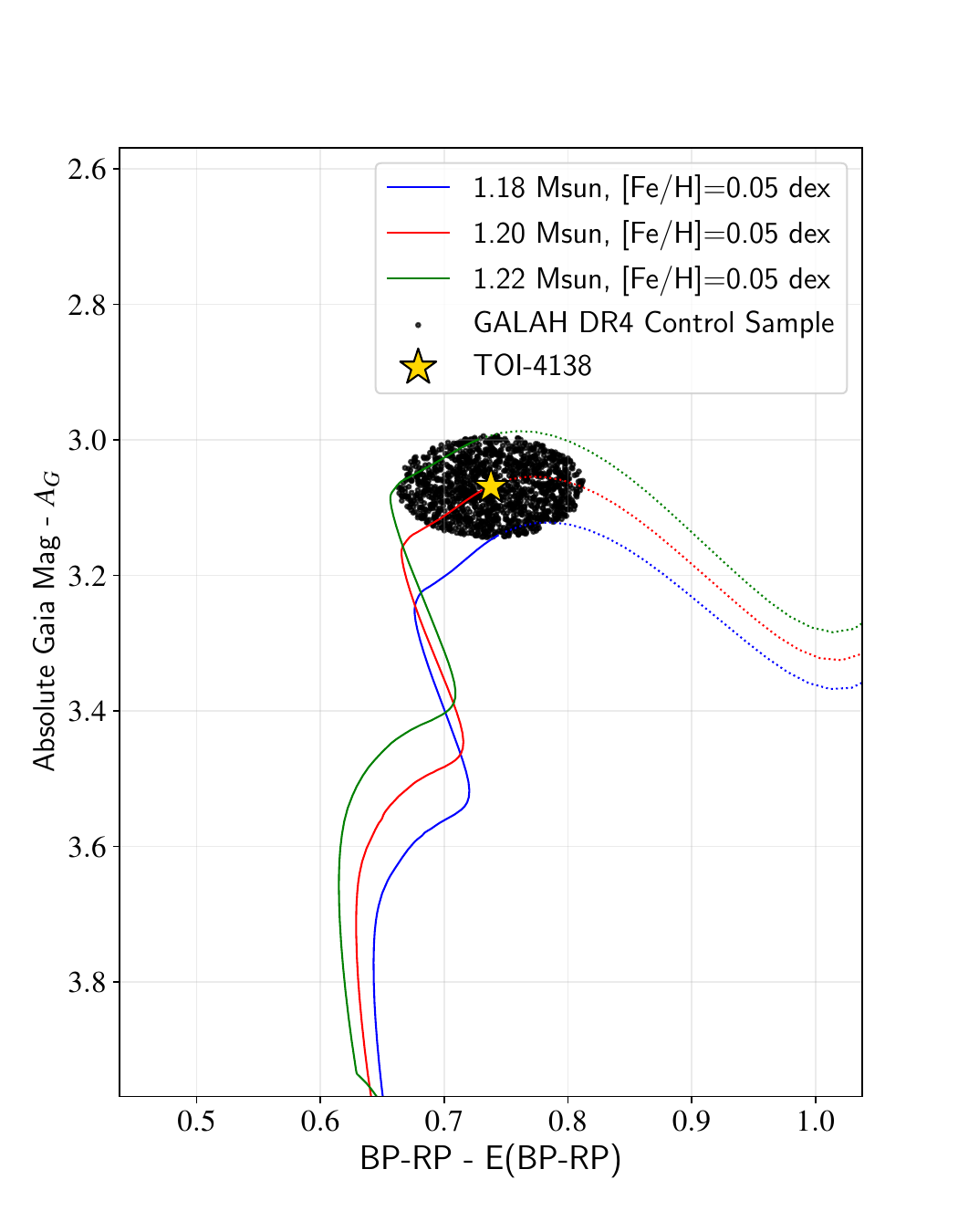}
    \caption{Color-magnitude diagram of the GALAH DR4 comparison sample of stars used to examine the significance of \confirmedtoi's Li anomaly. Three MIST evolutionary tracks are shown with metallicities equal to the median metallicity of \confirmedtoi. The solid lines represent the main sequence of each evolutionary track, while the dashed lines represent the subgiant branch. \confirmedtoi's median stellar mass is $1.188^{+0.11}_{-0.058}$ \msun.}
    \label{fig:galah_cmd_toi4138}
\end{figure*}

\begin{figure*}
    \centering   \includegraphics[width=0.7\textwidth,height=\textheight,keepaspectratio]{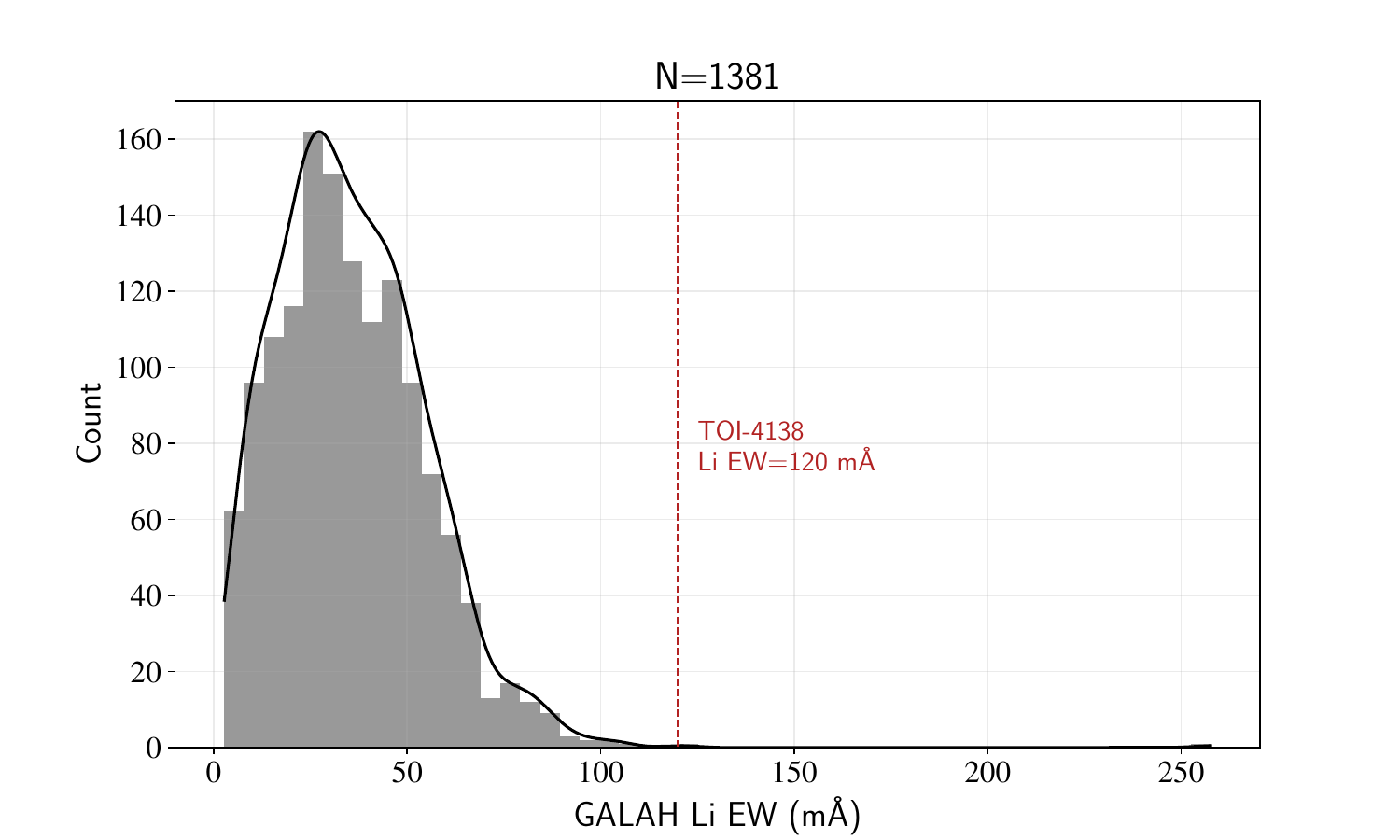}
    \caption{Histogram of lithium equivalent widths from our GALAH DR4 example. The solid curve is a kernel density estimation of the values represented in the histogram. The red dashed line represents the measured lithium equivalent width for \confirmedtoi of $121 \pm 12$ m\AA. The median equivalent width and median absolute deviation of the control sample are 33 m\AA\ and 13.3 m\AA, respectively. \confirmedtoi's Li enhancement is unusual for a star of its mass and age.}
    \label{fig:galah_hist_toi4138}
\end{figure*}

In order to place the measured strength of the Li signatures into context, we built a sample of comparison stars for each system using GALAH Data Release 4 (DR4) \citep{Buder:2024}. For each star, we constructed a color-magnitude diagram (CMD) using the extinction-corrected \textit{Gaia} $G$-band absolute magnitude and $G_{\textrm{BP}} - G_{\textrm{RP}}$ color. To ensure that the comparison stars are of a similar mass and age, we selected stars within a circle centered on the target star's color and absolute $G$ magnitude, with a radius of 0.075. This radius was selected as it is larger than each target's color and magnitude uncertainties and encompasses stars of a similar evolutionary stage within the target star's mass uncertainties. The control sample selection is illustrated in Figure \ref{fig:galah_cmd_toi4138}, using \confirmedtoi as an example. Then, we removed stars that failed to meet the following criteria: \textit{Gaia} renormalised unit weight error (RUWE) $< 1.4$ (removes unresolved binary stars), \texttt{flag\_red} $= 0$ (good SNR per pixel on red camera),  \texttt{flag\_sp} $= 0$ (good spectral fit), \texttt{flag\_fe\_h} $= 0$ (good metal abundance constraint), \texttt{snr\_px\_ccd3} $ > 30$ (minimum SNR of 30), and \texttt{flag\_a\_li} $< 4$ (good Li abundance constraint). Additional information on each of these flags can be found in the GALAH DR4 Table Schema\footnote{\url{https://www.galah-survey.org/dr4/table_schema/}} and the best practices guide for using the data\footnote{\url{https://www.galah-survey.org/dr4/using_the_data/}}. This yielded a control sample size of 1381 for \confirmedtoi, 177 for TOI-4773, 3848 for TOI-5350, and 3883 for TOI-6420. TOI-4773's relatively small sample size is a consequence of TOI-4773's higher mass and younger age, placing it in a region of the CMD with fewer stars. After comparing our targets to their respective GALAH control samples, we find that one of them, \confirmedtoi, stands out as a significant outlier, with a modified Z-score \citep{Iglewicz:1993} of 4.47. A histogram showing the Li EW of \confirmedtoi as it compares to the comparison sample is shown in Figure \ref{fig:galah_hist_toi4138}. This modified Z-score implies that \confirmedtoi's Li enhancement (Li EW = $120. \pm 13$ m\AA) is statistically significant and unusual for a star of its mass and age. \confirmedtoi is in the 99.86th percentile of its comparison GALAH DR4 control sample in terms of Li EW, with only two stars out of 1381 having a larger Li EW. TOI-4773 is moderately enhanced, in the 80th percentile of its control sample, but with a modified Z-score of 2.79, its Li enhancement is not statistically significant.

To further characterize the lithium content of each star, we used two independent codes, \texttt{pyMOOGi}\footnote{\url{https://github.com/madamow/pymoogi}} and \texttt{stardis}\footnote{\url{https://github.com/tardis-sn/stardis}}, to measure the Li abundance in each star. \texttt{pyMOOGi} is a Python wrapper for the FORTRAN code \texttt{MOOG}, which is a visual tool that does spectral line analysis and spectral synthesis. When measuring the lithium abundances with \texttt{PyMOOGi}, we used interpolated ATLAS9 model atmospheres \citep{Meszaros:2012} that were retrieved using \texttt{PyKMOD}\footnote{\url{https://github.com/kolecki4/PyKMOD}}. \texttt{stardis}, on the other hand, is a newly developed native Python stellar spectral synthesis code \citep{Shields:2025} that we attach to a simple residual minimization framework to compare to our observed spectra and constrain abundances and their uncertainties. We opted to use a different set of model atmospheres from the MARCS model atmosphere grid \citep{Gustafsson:2008} and a newer solar abundance source \citep{Asplund:2021}. The measured Li abundances, A(Li), are reported in Table \ref{tab:lithium}. While GALAH DR4 does include Li abundances as one of its data products, the majority of stars have too little Li for their abundances to be measured, so therefore the Li abundances in the GALAH sample only represent the stars with enhanced Li. Additionally, the choices made in measuring Li abundances can lead to large variations in reported abundances; e.g., A(Li) in one of the most cited source of solar abundances \citep{Asplund:2009} is $\sim 10\%$ different than A(Li) from their more recent work \citep{Asplund:2021}. For these reasons, it is more appropriate to compare the EW of our target stars to their respective GALAH control samples.

\section{Discussion}\label{sec:discussion}

In this article, we reanalyze a benchmark HJ system, \confirmedtoi, and discover four additional systems that occupy sparse regions of the exoplanet mass-radius diagram. We discuss the implications of the Li detections in four of the host stars and review the properties that make each planet and host star unique. Uncommon planets such as these provide useful tests to models of planet formation and measured Li abundances provide independent constraints on the age and evolutionary history of each system. Each planet is also an important contribution to the self-consistent sample of HJs being built by the MEEP survey.

\subsection{Interpretations of Lithium Detections}\label{subsec:lithium_interp}

We used the Python package \texttt{EAGLES}\footnote{\url{https://github.com/robdjeff/eagles}} \citep{Jeffries:2023} to estimate the age of the three host stars that are cooler than 6500\,K and have a detectable Li feature: \confirmedtoi, TOI-5350, and TOI-6420. \texttt{EAGLES} predicted a broad probability distribution for TOI-5350, using only the stellar effective temperature and Li EW as constraints: $2.0^{+4.0}_{-1.9}$\,Gyr. This agrees within $1\sigma$ with the \exofast age, $3.1^{+2}_{-1.4}$\,Gyr. Because of TOI-6420's uncertain Li EW, \texttt{EAGLES} can only place a 95\% lower limit on its age at 200\,Myr, which is also consistent with the age estimate from \exofast: $4.2^{+2.6}_{-1.9}$\,Gyr.

\confirmedtoi, however, owing to its strong Li feature (Li EW$ = 120. \pm 13$ m\AA), has an expected age of $103^{+171}_{-95}$\,Myr, which is inconsistent with the median age from the \exofast fit, $5.6^{+1.2}_{-1.7}$\,Gyr. In order to investigate this discrepancy, we searched for additional signs of youth. One compelling sign of youth is membership to a comoving association of stars with a known age. Using the Python package \texttt{Comove}\footnote{\url{https://github.com/adamkraus/Comove}} \citep{Tofflemire:2021}, we searched for nearby comoving companions and found none within a 25 pc radius around \confirmedtoi. \confirmedtoi Additionally, we found no evidence of infrared excess, excessively fast stellar rotation ($v\sin{i_\star} = 5.45 \pm 0.074$ km s$^{-1}$), or enhanced X-ray emission, all of which would be indicators of youth. Lastly, we employed the \texttt{Sagitta} tool\footnote{\url{https://github.com/hutchresearch/Sagitta}} \citep{McBride:2021} to measure \confirmedtoi's pre-main sequence probability using deep neural networks trained on \textit{Gaia} and 2MASS photometry. \texttt{Sagitta} predicted a low pre-main sequence probability of 0.005\% implying that youth is an unlikely cause for the Li enhancement. Instead, we choose to adopt the age constraints from the \exofast fit which are in better agreement with the results from \cite{Montalto:2022}. Therefore, the primordial Li in \confirmedtoi's photosphere should have been mixed into its interior and destroyed during the star's pre-main sequence and main sequence lifetime. As a subgiant star, \confirmedtoi is also too young for its Li to be self-enriched by the Cameron-Fowler conveyor \citep{Cameron:1971}, as this self-enrichment becomes important only after the star has turned onto the red giant branch from the subgiant branch.

Since the available data imply that \confirmedtoi is not young and that its enhanced Li is not primordial, we can explore a couple of other viable mechanisms to enrich the star with Li. The most obvious mechanism is the ingestion of Li-rich planetary material. Planetary engulfment events are likely to be somewhat common, as roughly 1\% of FGK main-sequence stars host HJs \citep{Wright:2010} and it is expected that many of these HJs are engulfed as the host star evolves off of the main sequence \citep{Schlaufman:2013, Villaver:2014}. Evidence of planetary engulfment is harder to uncover, as the infrared transient is short-lived \citep{De:2023} and the expected lithium enhancement may not be detectable or may be short-lived, depending on host star parameters \citep{Soares-Furtado:2021}. \confirmedtoi, however, occupies a region of parameter space where \cite{Soares-Furtado:2021} predicts that if a gas giant were to be ingested, the engulfing star would show a statistically significant Li enhancement for as long as 1.5\,Gyr. \cite{Behmard:2023} also argues that solar metallicity stars with masses $\approx 1.1 - 1.2$ \msol have the largest and longest-lived chemical signatures from planet engulfment. Therefore, \confirmedtoi is a compelling planetary engulfment candidate and an excellent laboratory to investigate the final stages of planetary evolution. Simulating the dynamics of this system could elucidate the circumstances leading to the potential engulfment event, which could have been the product of two scenarios: the surviving planet, \confirmedtoi b, underwent large-scale migration, leading to the orbits of nearby planets changing and one or more planets being engulfed, or that \confirmedtoi b migrated inwards along with other planet(s), which were engulfed by the host star as it evolved off of the main sequence and expanded.

Another plausible explanation is that the Li in \confirmedtoi's photosphere comes from pollution caused by a nearby classical nova eruption, the ejecta of which are known to be Li-rich \citep{Starrfield:1978}. These events are relatively common, as \cite{Kawash:2022} found that $26 \pm 5$ novae likely erupt in the Milky Way each year. However, this is a possibility that is difficult to confirm or reject because of the long Li survival time. Using SIMBAD\footnote{\url{https://simbad.cds.unistra.fr/simbad/}} and the American Association of Variable Star Observers (AAVSO) Variable Star Index\footnote{\url{https://www.aavso.org/vsx/}}, we performed a cone search around \confirmedtoi with a radius of one degree and found no confirmed nova eruptions or nova remnants. This does not rule out nova pollution as the source of the Li, but further investigation is beyond the scope of this paper.

\subsection{\confirmedtoi b: Reanalysis of an Inflated Hot Jupiter Orbiting an Evolved Star}\label{subsec:toi4138}

\begin{figure*}
    \centering
    \includegraphics[width=0.7\textwidth,height=\textheight,keepaspectratio]{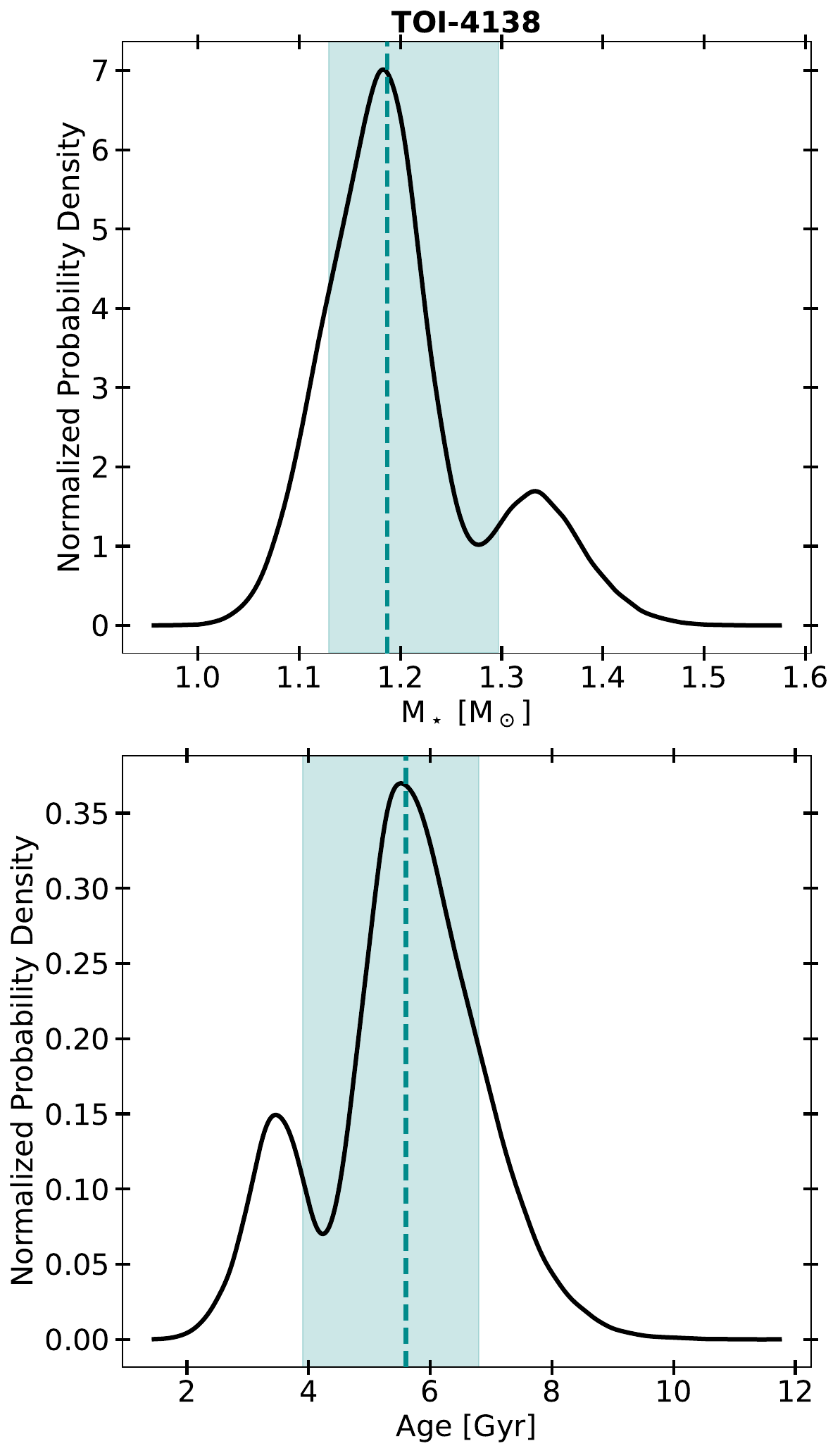}
    \caption{Gaussian kernel density estimation of \confirmedtoi's bimodal mass and age posterior distributions. In both cases, the probability density has been normalized to the integral of the curve. The cyan dashed lines represent the reported median mass and age from Table \ref{tab:median}, while the cyan shaded regions represent the reported 68\% confidence regions. The less massive solution is significantly favored.}
    \label{fig:pdf_toi4138}
\end{figure*}

\cite{Montalto:2022} announced the discovery of \confirmedtoi b (referred to as TIC 257060897b in their work) and described it as an extremely low-density hot Jupiter orbiting a metal-rich, rapidly evolving, subgiant star. To reanalyze the system and incorporate it within the self-consistent sample, we treated it as if it were an unconfirmed planet and obtained ground-based follow-up photometry, radial velocities, and speckle imaging of the system. We then used these data to globally fit the system using \exofast. The resulting solution is bimodal in both stellar mass and age, as is often the case when a star is near an evolutionary transition point such as the main sequence turnoff. To better characterize each mode of the distribution, we split the solution at a mass of 1.27 \msol, the local minimum between the two distribution peaks. The solution with a smaller host star mass ($1.174^{+0.043}_{-0.052}$ \msol) and larger age ($5.87^{+1.1}_{-0.78}$ Gyr) is favored with an 80.5\% probability. We present the two separate resulting solutions in Table \ref{tab:bimodal}, comparing them against the results from \cite{Montalto:2022}, and in Figure \ref{fig:pdf_toi4138}.

\providecommand{\bjdtdb}{\ensuremath{\rm {BJD_{TDB}}}}
\providecommand{\feh}{\ensuremath{\left[{\rm Fe}/{\rm H}\right]}}
\providecommand{\teff}{\ensuremath{T_{\rm eff}}}
\providecommand{\teq}{\ensuremath{T_{\rm eq}}}
\providecommand{\ecosw}{\ensuremath{e\cos{\omega_*}}}
\providecommand{\esinw}{\ensuremath{e\sin{\omega_*}}}
\providecommand\msun{M$_\odot$\xspace}
\providecommand{\rsun}{R$_\odot$\xspace}
\providecommand{\lsun}{L$_\odot$\xspace}
\providecommand{\mj}{\ensuremath{\,M_{\rm J}}}
\providecommand{\rj}{\ensuremath{\,R_{\rm J}}}
\providecommand{\me}{\ensuremath{\,M_{\rm E}}}
\providecommand{\re}{\ensuremath{\,R_{\rm E}}}
\providecommand{\fave}{\langle F \rangle}
\providecommand{\fluxcgs}{10$^9$ erg s$^{-1}$ cm$^{-2}$}
\providecommand{\tess}{\textit{TESS}\xspace}
\begin{table*}
\centering
\caption{Median values and 68\% confidence intervals for TOI-4138's bimodal solution, compared to the discovery work by \citet{Montalto:2022}.}
\label{tab:bimodal}
\scriptsize
\begin{tabular}{l l c c c}
\hline
 & & High-mass solution & Low-mass solution & \citet{Montalto:2022}\\
 & & (19.5\% probability) & (80.5\% probability) & \\
\hline
\multicolumn{5}{l}{\textbf{Stellar Parameters}:} \\
$M_*$ & Mass (\msun) & $1.338^{+0.048}_{-0.04}$ & $1.174^{+0.043}_{-0.052}$ & $1.32 \pm 0.04$ \\
$R_*$ & Radius (\rsun) & $1.849^{+0.06}_{-0.058}$ & $1.865^{+0.049}_{-0.052}$ & $1.82 \pm 0.05$ \\
$L_*$ & Luminosity (\lsun) & $4.05^{+0.15}_{-0.14}$ & $4.03 \pm 0.15$ & $4.07 \pm 1.05$ \\
$\rho_*$ & Density (cgs) & $0.299^{+0.029}_{-0.026}$ & $0.254^{+0.023}_{-0.019}$ & $0.22 \pm 0.01$ \\
$\log{g}$ & Surface gravity (cgs) & $4.031^{+0.028}_{-0.026}$ & $3.965^{+0.026}_{-0.025}$ & $4.2 \pm 0.1$ \\
$T_{\rm eff}$ & Effective temperature (K) & $6019.0^{+94}_{-91}$ & $5987.0^{+86}_{-81}$ & $6128 \pm 57$ \\
$[{\rm Fe/H}]$ & Metallicity (dex) & $0.087^{+0.078}_{-0.077}$ & $0.048^{+0.076}_{-0.068}$ & $0.20 \pm 0.04$ \\
$[{\rm Fe/H}]_{0}$ & Initial metallicity & $0.145^{+0.07}_{-0.066}$ & $0.089^{+0.068}_{-0.065}$ & --- \\
Age & Age (Gyr) & $3.45^{+0.47}_{-0.5}$ & $5.87^{+1.1}_{-0.78}$ & $3.47 \pm 1.10$ \\
EEP & Equivalent evolutionary phase & $406.3^{+8.4}_{-13}$ & $452.8^{+4.4}_{-5.3}$ & --- \\
$A_V$ & V-band extinction (mag) & $0.066^{+0.018}_{-0.032}$ & $0.06^{+0.022}_{-0.034}$ & $0.08 \pm 0.02$ \\
$d$ & Distance (pc) & $497.2 \pm 3.7$ & $496.8 \pm 3.7$ & $498 \pm 13$ \\
\multicolumn{5}{l}{\textbf{Planetary Parameters}:} \\
$P$ & Period (days) & $3.6600364^{+0.00000099}_{-0.00000098}$ & $3.66003639 \pm 0.00000098$ & $3.660028 \pm 0.000006$\\
$R_{\rm P}$ & Radius (\rj) & $1.528^{+0.049}_{-0.048}$ & $1.542^{+0.042}_{-0.044}$ & $1.49 \pm 0.04$ \\
$M_{\rm P}$ & Mass (\mj) & $0.684^{+0.032}_{-0.03}$ & $0.626 \pm 0.03$ & $0.67 \pm 0.03$\\
$T_C$ & Time of conjunction (\bjdtdb) & $2459660.60713 \pm 0.00017$ & $2459660.60712 \pm 0.00017$ & $ 2458708.9983 \pm 0.0003$ \\
$T_0$ & Optimal conjunction time (\bjdtdb) & $2459473.94542 \pm 0.00013$ & $2459473.94542 \pm 0.00013$ & --- \\
$a$ & Semi-major axis (AU) & $0.05122^{+0.0006}_{-0.00052}$ & $0.04904^{+0.00059}_{-0.00074}$ & $ 0.051 \pm 0.002 $ \\
$i$ & Inclination (Degrees) & $85.05^{+0.32}_{-0.33}$ & $84.46^{+0.35}_{-0.36}$ & $ 86.0 \pm 0.7 $ \\
$e$ & Eccentricity & $0.079 \pm 0.037$ & $0.034^{+0.028}_{-0.02}$ & $ 0.03 \pm 0.02 $ \\
$\omega_*$ & Argument of periastron (Degrees) & $-66.0^{+17}_{-9.7}$ & $-40.0^{+63}_{-29}$ & $ 20 \pm 72 $ \\
$\teq$ & Equilibrium temperature (K) & $1743.0 \pm 17$ & $1780.0^{+19}_{-18}$ & $ 1762 \pm 21 $ \\
$\tau_{\rm circ}$ & Tidal circularization timescale (Gyr) & $0.0768^{+0.0092}_{-0.0085}$ & $0.0658^{+0.0088}_{-0.008}$ & --- \\
$K$ & RV semi-amplitude (m/s) & $74.1^{+3}_{-2.9}$ & $74.1 \pm 2.9$ & $ 74 \pm 3 $ \\
$R_{\rm P}/R_*$ & Radius of planet in stellar radii & $0.08492^{+0.00036}_{-0.00037}$ & $0.085^{+0.00033}_{-0.00034}$ & $ 0.0841 \pm 0.0009 $ \\
$a/R_*$ & Semi-major axis in stellar radii & $5.96^{+0.19}_{-0.18}$ & $5.64^{+0.16}_{-0.14}$ & $ 6.05 \pm 0.09 $ \\
Depth & \tess flux decrement at mid-transit & $0.007846^{+0.000046}_{-0.000044}$ & $0.007851^{+0.000045}_{-0.000044}$ & --- \\
$\tau$ & Ingress/egress transit duration (days) & $0.02169^{+0.00094}_{-0.00093}$ & $0.02192^{+0.00088}_{-0.00087}$ & --- \\
$T_{14}$ & Total transit duration (days) & $0.19739^{+0.00083}_{-0.0008}$ & $0.1976^{+0.0008}_{-0.00079}$ & $ 0.194 \pm 0.005 $ \\
$b$ & Transit impact parameter & $0.552^{+0.024}_{-0.027}$ & $0.557^{+0.022}_{-0.024}$ & $0.42 \pm 0.08$ \\
$\rho_{\rm P}$ & Density (cgs) & $0.238^{+0.025}_{-0.023}$ & $0.212^{+0.02}_{-0.018}$ & $0.25 \pm 0.02$ \\
$\log{g_{\rm P}}$ & Surface gravity (cgs) & $2.86^{+0.032}_{-0.031}$ & $2.815^{+0.03}_{-0.029}$ & $2.87 \pm 0.03$ \\
$M_{\rm P}/M_*$ & Mass ratio  & $0.000487 \pm 0.00002$ & $0.000511 \pm 0.000021$ & --- \\
$d/R_*$ & Separation at mid-transit  & $6.37^{+0.42}_{-0.4}$ & $5.73^{+0.36}_{-0.25}$ & --- \\
\hline
\end{tabular}
\end{table*}

We find that our disfavored high-mass solution for \confirmedtoi agrees well within $\sim 1\sigma$ with most of the reported parameters from \cite{Montalto:2022}. However, the solution that our \exofast global fit favors disagrees with the \cite{Montalto:2022} solution; in particular, the stellar mass and several derived parameters dependent on the stellar mass disagree by $\sim 2-3\sigma$. One of the possible reasons for this is the different stellar evolution models used: \cite{Montalto:2022} used PARSEC \citep{Bressan:2012}, while we used MIST \citep{Paxton:2011}. To test this and see how the stellar mass posterior distribution changes depending on the stellar evolution model used, we re-ran the \exofast fit for \confirmedtoi twice, using PARSEC and Yonsei-Yale \citep{Yi:2001} models instead of MIST. We find that each of the \exofast fits agree that the lower mass solution is favored. Instead, one of the primary reasons for this is likely because different priors were adopted and different fitting software were used. In particular, the metallicity prior used by \cite{Montalto:2022}, [Fe/H] = $+0.20 \pm 0.04$, is larger than the spectroscopic metallicity prior we adopt of [Fe/H] = $0.062 \pm 0.094$.

Both of our \exofast solutions agree with the determination made by \cite{Montalto:2022} that \confirmedtoi b has an inflated radius. In fact, the favored \exofast solution predicts an even smaller density than \cite{Montalto:2022}, placing \confirmedtoi b firmly in the 6th percentile of all measured planet densities\footnote{NASA Exoplanet Archive accessed 2025 June 11.}. \cite{Montalto:2022} argues that \confirmedtoi b may have been re-inflated as the luminosity of the host star quickly increased as it evolved off of the main sequence.

Future observations of \confirmedtoi may also yield constraints on a possible second companion orbiting \confirmedtoi on a longer period. As discussed in \S \ref{subsubsec:tres}, we opted to allow for a zero-point offset between the two seasons of TRES RVs to better characterize \confirmedtoi b. If future RVs reveal a turnover in the RVs, it would be possible to characterize both the period and mass of an additional companion. Understanding the orbital properties of additional companions could provide further context surrounding the migration of \confirmedtoi b and the possibility that engulfment of an inner planet explains the observed Li enrichment.

\subsection{Four Super-Jupiters Orbiting FGK Stars}\label{subsec:superjupiters}

\begin{figure*}
    \centering
    \includegraphics[width=0.7\textwidth,keepaspectratio]{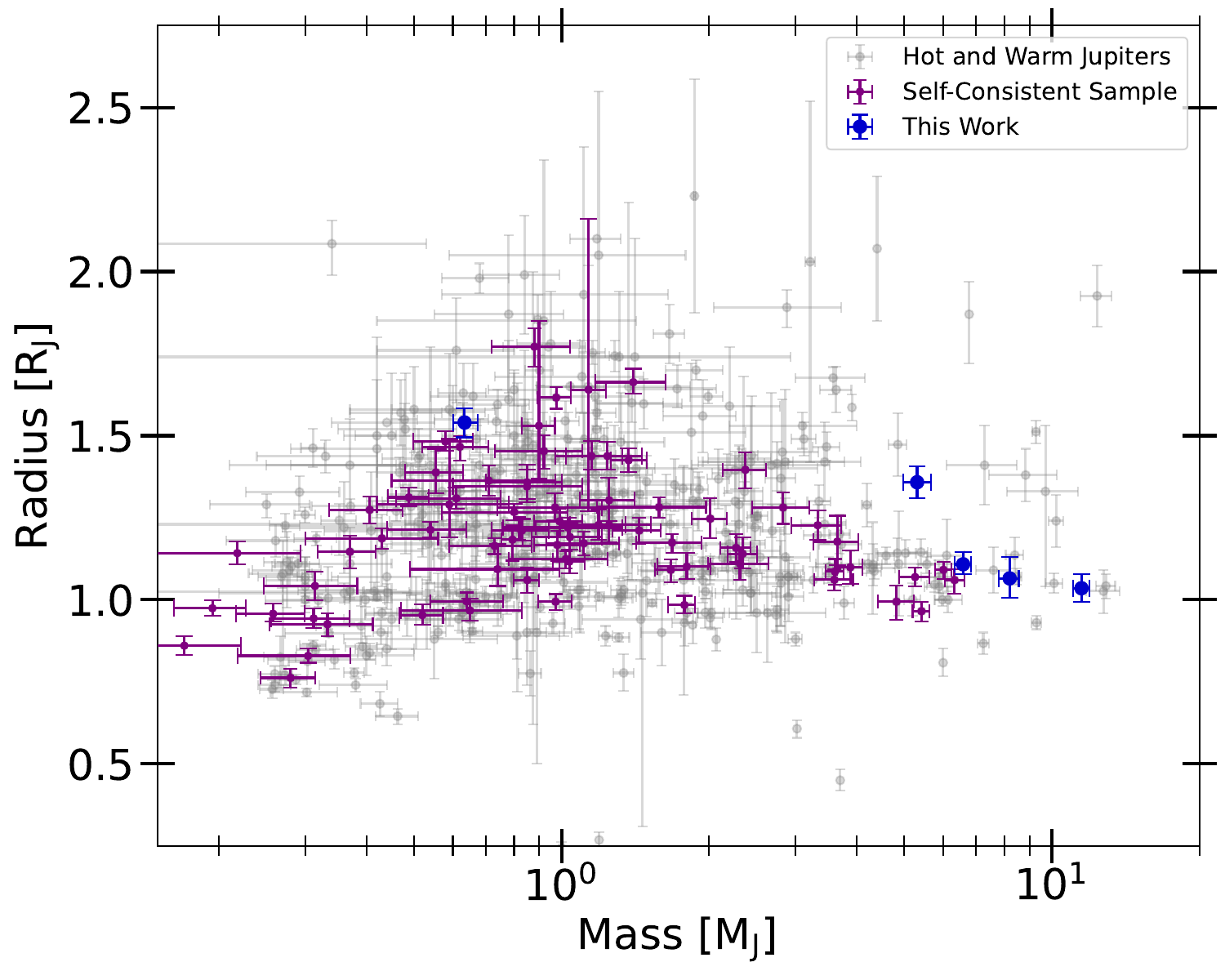}
    \caption{Mass vs. radius diagram of the existing sample of confirmed hot and warm Jupiters, colored in grey (retrieved through the NASA Exoplanet Archive on \NEAdate), compared to the systems analyzed in the way described in this article (\citealp{Rodriguez:2021, Ikwut-Ukwa:2022, Yee:2022, Yee:2023a, Rodriguez:2023, Schulte:2024, RodriguezMartinez:2025, Yee:2025}), colored in purple. We adopt the definition for a hot and warm Jupiter as prescribed by \citealp{Dawson:2018}, M$_{\mathrm{p}} > 0.25$ \mj and P $< 100$ days. The systems in this work are colored in blue and include some of the most massive giant exoplanets in our homogeneous sample.}
    \label{fig:m_r}
\end{figure*}

In addition to \confirmedtoi, we discovered and characterized four additional planets in this article: TOI-4773 b, TOI-5261 b, TOI-5350 b, and TOI-6420 b. All four of these objects are more massive than 5 \mj, occupying a sparse region of the giant planet mass-radius diagram, as shown in Figure \ref{fig:m_r}. TOI-4773 b is an inflated super-Jupiter orbiting a young main sequence star above the Kraft break. As a consequence of the star's only moderately fast rotation ($v\sin{i_\star} = 15.3 \pm 0.29$ kms$^{-1}$), this system is one of the hottest with a measured planetary eccentricity. TOI-5261 b is an eccentric, $11.49^{+0.43}_{-0.46}$ \mj super-Jupiter orbiting a metal-rich Solar analogue. Its mass is large enough that the deuterium burning limit of 13 \mj, the traditionally accepted lower mass limit for brown dwarfs, is only more massive by $3.5\sigma$. While there is no analog to this in the Solar System, the host star has quite similar properties to the Sun: its mass, radius, age, and luminosity are within 1$\sigma$ of the Sun. Therefore, this system may act as a laboratory to understand planetary system formation and evolution in a context that can be easily compared to the Solar System and address the question of why a planet like TOI-5261 b formed in this system but not in the Solar System. Finally, TOI-5261 b has significant nonzero orbital eccentricity ($e = 0.1585^{+0.01}_{-0.0078}$), which is likely a remnant from its migration to its current orbit.

\begin{figure*}
    \centering
    \includegraphics[width=\textwidth,height=0.8\textheight,keepaspectratio]{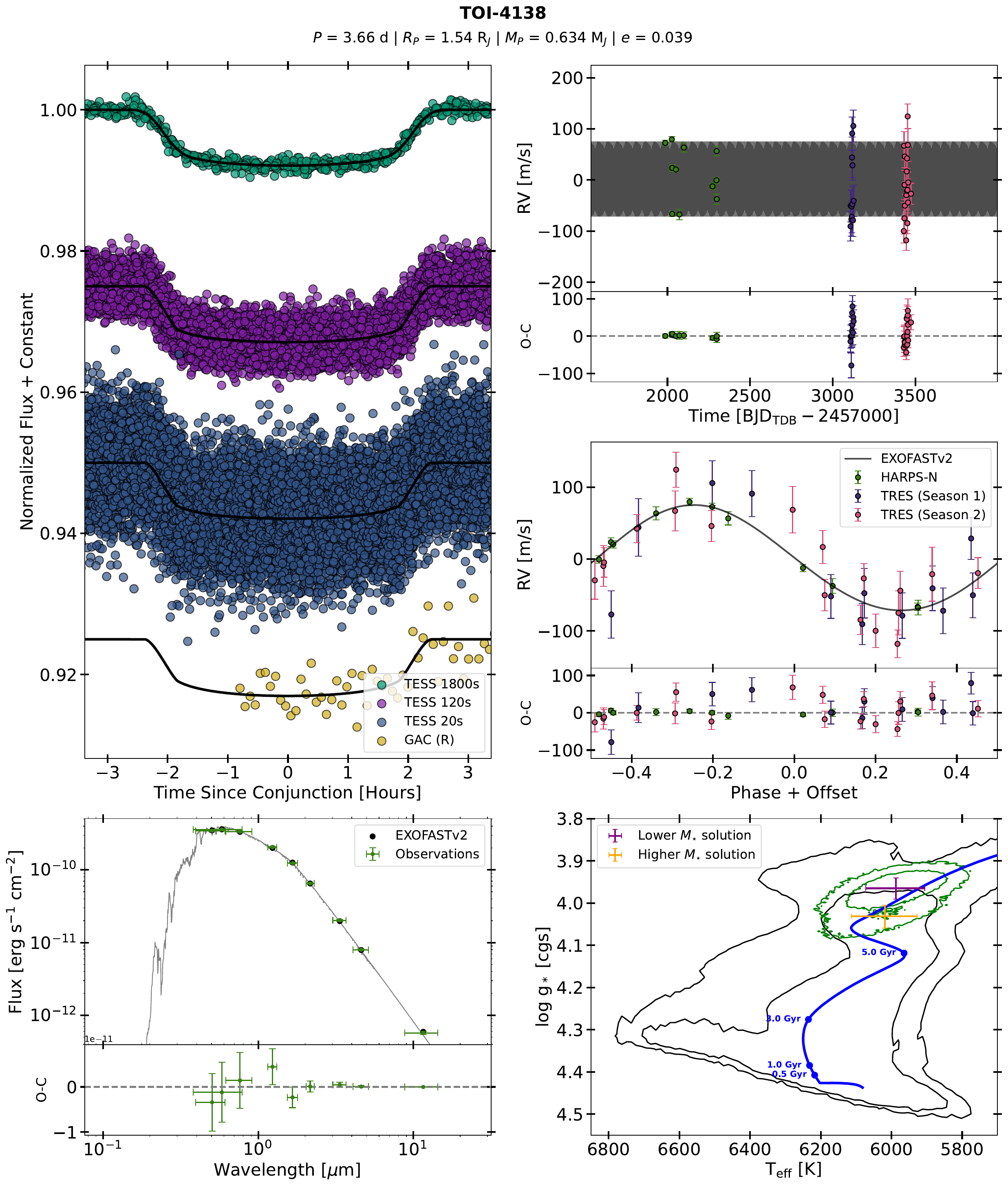}
    \caption{Photometric and radial velocity observations of the \confirmedtoi system.
    \textbf{Upper left:} Phase-folded, unbinned, transits of \confirmedtoi b shown in comparison to the best-fit time of conjunction with an arbitrary normalized flux offset. Multiple \tess sectors in the same cadence are stacked on top of each other.
    \textbf{Bottom left:} The spectral energy distribution of the target star compared to the best-fit \exofast model. Residuals are shown on a linear scale, using the same units as the primary y-axis.
    \textbf{Upper right:} RV observations versus time, including any significant long-term trend. The residuals are shown in the subpanel below in the same units.
    \textbf{Middle right:} RV observations phase-folded using the best-fit ephemeris from the \exofast global fit. The phase is shifted so that the transit occurs at Phase + Offset = 0. The residuals are shown in the subpanel below in the same units.
    \textbf{Bottom right:} The evolutionary track and current evolutionary stage of the planet according to the best-fit MESA Isochrones and Stellar Tracks (MIST) model. The blue line indicates the best-fit MIST track, while the black contours show the 1$\sigma$ and 2$\sigma$ constraints on the star's current \teff and log g from the MIST isochrone alone. The green contours represent the 1$\sigma$ and 2$\sigma$ constraints on the star's \teff and log g from the \exofast global fit, combining constraints from observations of the star and planet. The purple and orange crosses indicate the medians and 68\% confidence intervals for each \exofast solution reported in Table \ref{tab:bimodal}.} % split pdf at Mstar = 1.27 Msol
    \label{fig:toi4138}
\end{figure*}

\begin{figure*}
    \centering    \includegraphics[width=\textwidth,height=\textheight,keepaspectratio]{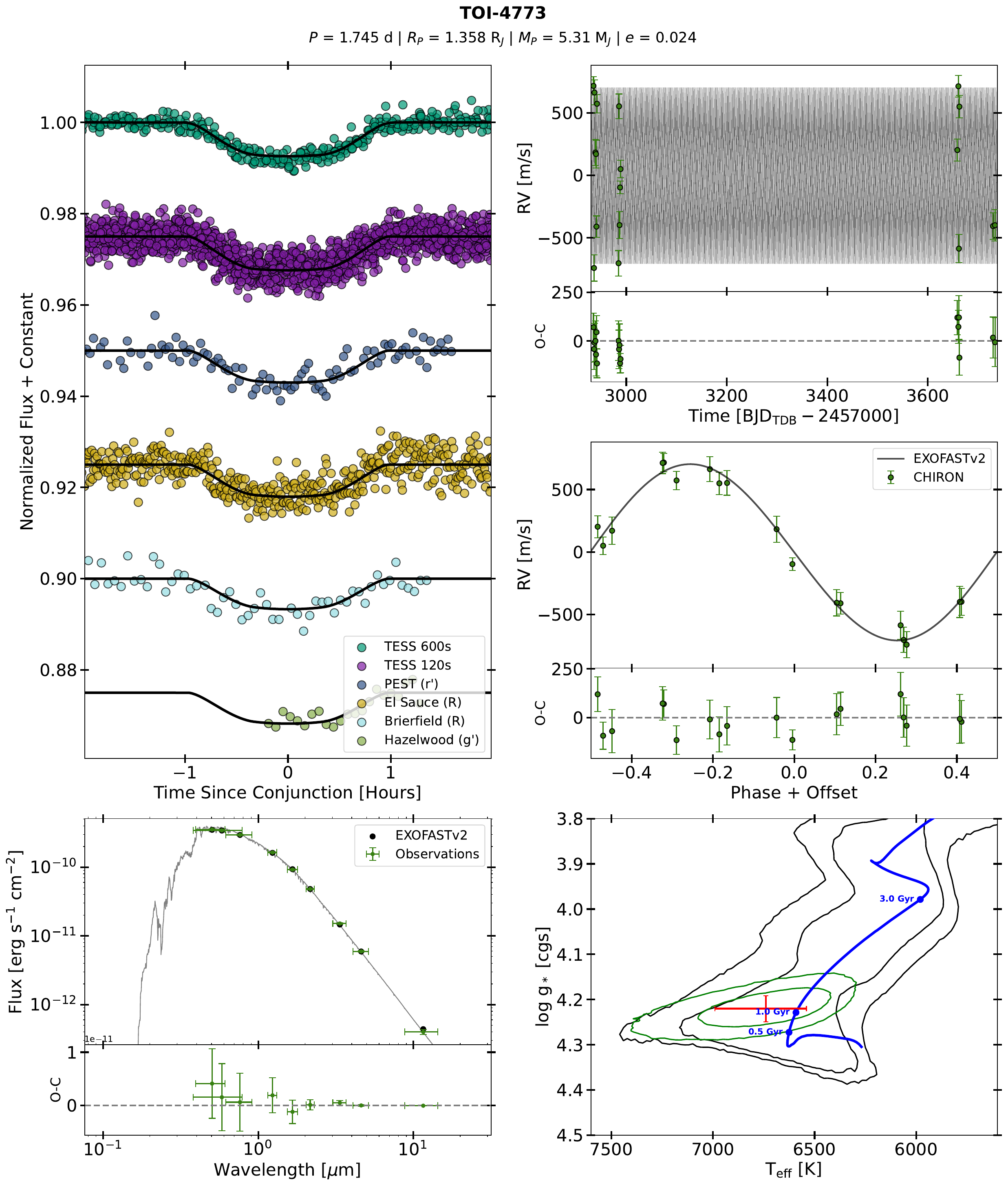}
    \caption{Same as Figure \ref{fig:toi4138}, but for TOI-4773. The red cross in the bottom right plot indicates the median and 68\% confidence interval reported in Table \ref{tab:median}.}
    \label{fig:toi4773}
\end{figure*}

\begin{figure*}
    \centering    \includegraphics[width=\textwidth,height=\textheight,keepaspectratio]{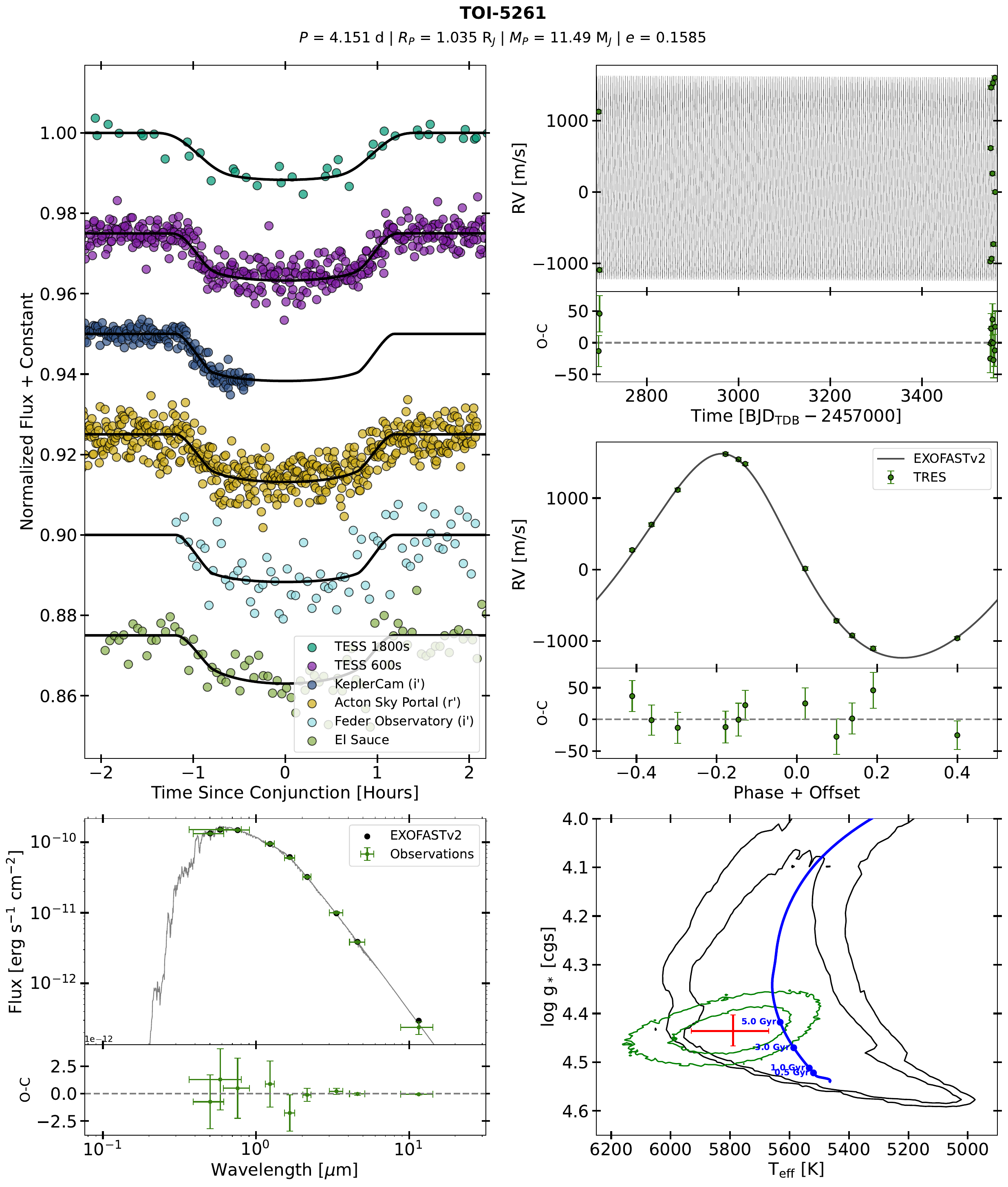}
    \caption{Same as Figure \ref{fig:toi4138}, but for TOI-5261. The red cross in the bottom right plot indicates the median and 68\% confidence interval reported in Table \ref{tab:median}.}
    \label{fig:toi5261}
\end{figure*}

\begin{figure*}
    \centering     \includegraphics[width=\textwidth,height=\textheight,keepaspectratio]{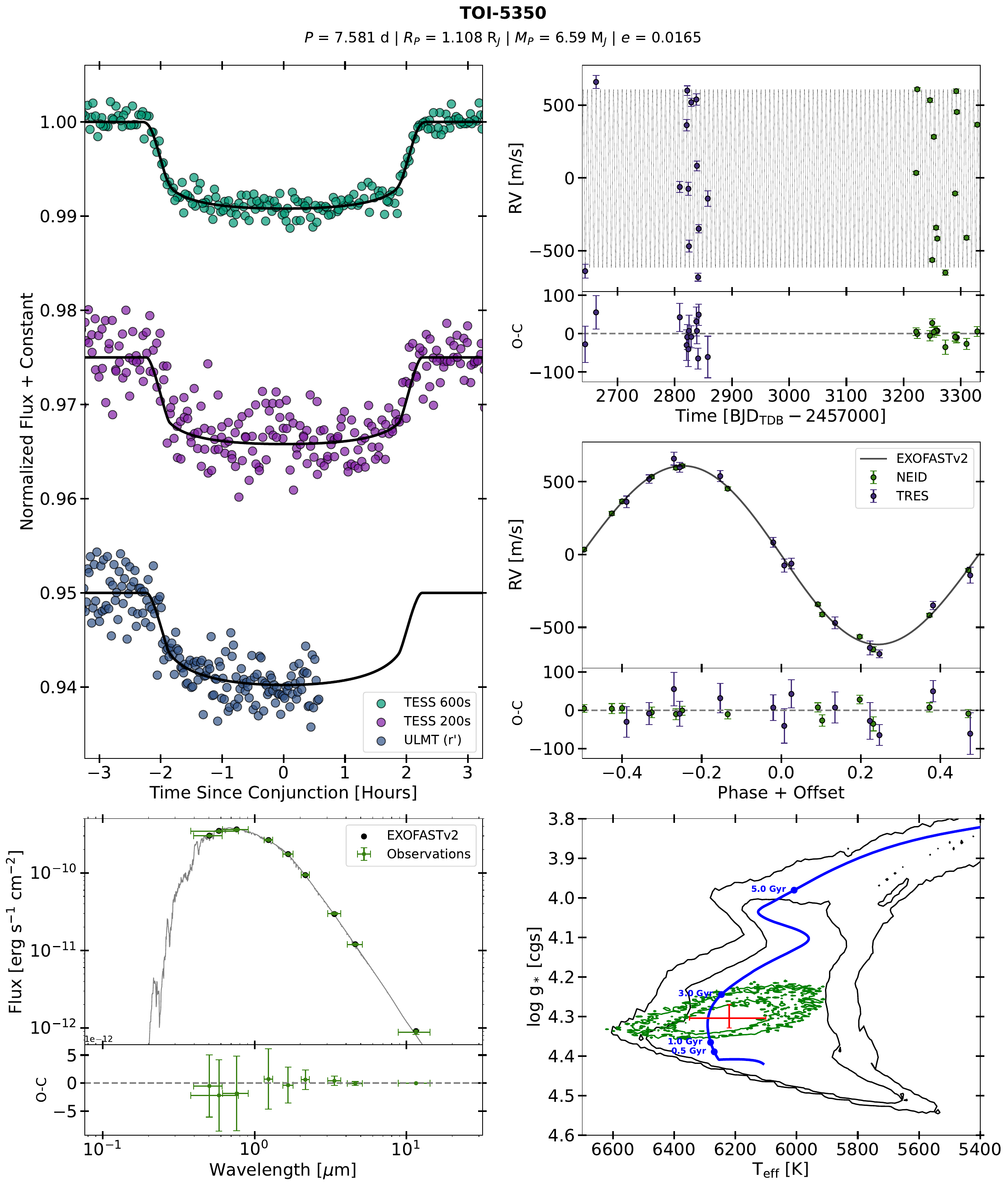}
    \caption{Same as Figure \ref{fig:toi4138}, but for TOI-5350. The red cross in the bottom right plot indicates the median and 68\% confidence interval reported in Table \ref{tab:median}.}
    \label{fig:toi5350}
\end{figure*}

\begin{figure*}
    \centering
    \includegraphics[width=\textwidth,height=\textheight,keepaspectratio]{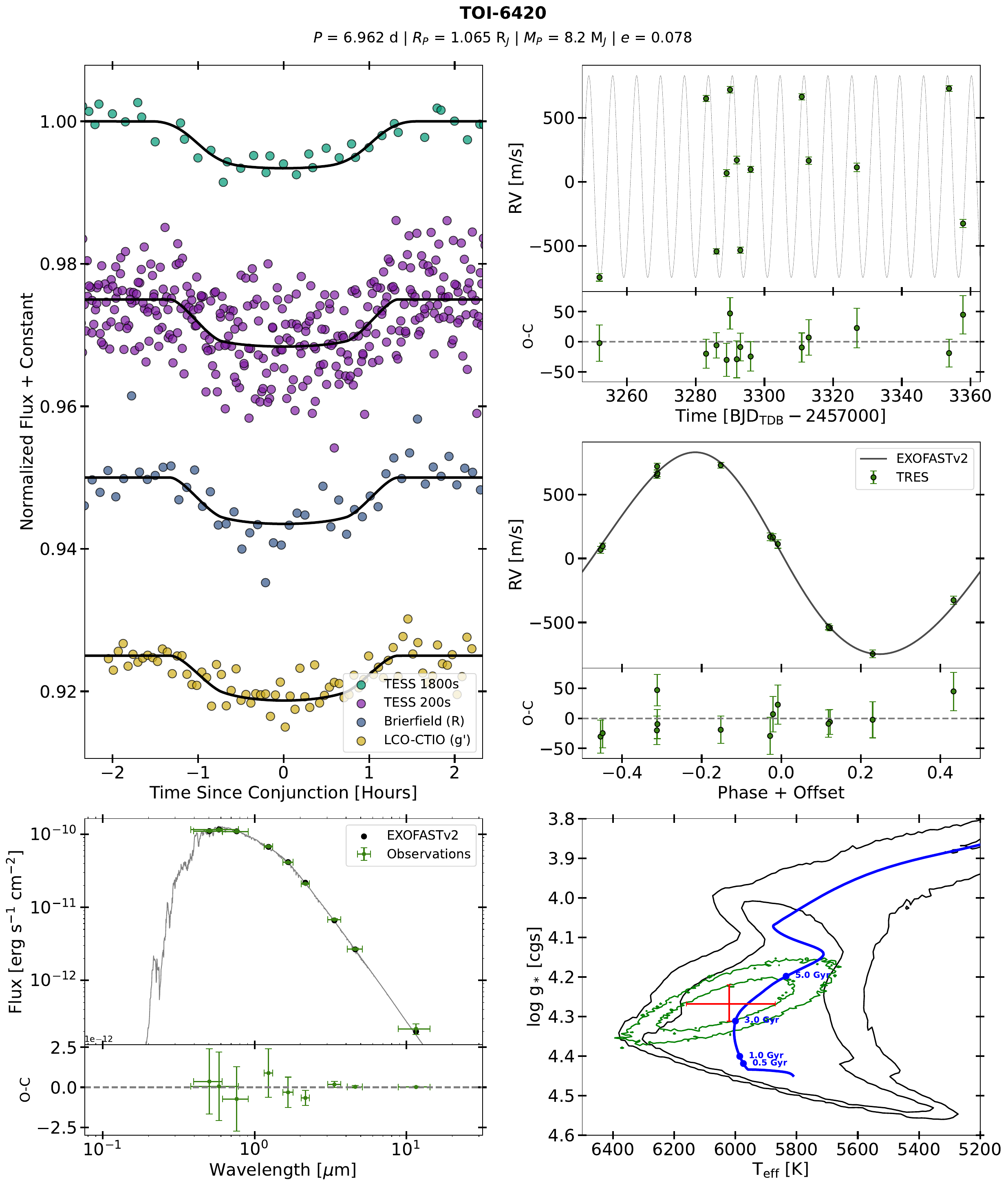}
    \caption{Same as Figure \ref{fig:toi4138}, but for TOI-6420. The red cross in the bottom right plot indicates the median and 68\% confidence interval reported in Table \ref{tab:median}.}
    \label{fig:toi6420}
\end{figure*}

The other two objects we have confirmed, TOI-5350 b and TOI-6420 b, are both massive HJs orbiting F-type stars. While the host star and planetary properties of these two systems are generally similar, TOI-6420's host star is metal-rich ([Fe/H] = $0.24^{+0.073}_{-0.077}$), unlike TOI-5350's roughly solar metal abundance. Each of the five systems presented in this article will be valuable systems in the final magnitude-limited, complete, self-consistent sample of HJs being constructed as part of the MEEP survey. A summary of each planetary system's key parameters and observations are presented in Figures \ref{fig:toi4138} - \ref{fig:toi6420}.

% come up with a better ending to this section that seamlessly flows into the next section?

\subsection{The Growing Self-Consistent Sample of Hot Jupiters}\label{subsec:sample}

As of \NEAdate, there were \totHJs{} HJs discovered via radial velocity and transit surveys. Of these, \transitingHJs{} transit and have constrained radii. However, the selection function for this sample is poorly defined and the analysis for these systems was done using a variety of techniques. Going forward, the MEEP survey \citep{Schulte:2024} aims to discover and characterize HJ systems with the same methodology to ensure self-consistency as a magnitude-limited sample of transiting HJs is completed. This paper adds five systems to the self-consistent sample of HJs orbiting FGK stars, which now numbers \selfconsistentHJs{} (\citealt{Rodriguez:2019, Rodriguez:2021, Ikwut-Ukwa:2022, Yee:2022, Yee:2023a, Rodriguez:2023, Schulte:2024, Yee:2025}). Once the sample is complete, it will be possible to inspect the statistics of the sample and address a large number of questions and tentative trends reported in the literature, such as the occurrence of HJs around FGK dwarfs and the frequency of disk migration and high-eccentricity tidal migration. The answers to these questions will enable a holistic understanding of transiting giant planets around the stars most similar to the Sun.

\begin{figure*}
    \centering
    \includegraphics[width=0.7\textwidth,height=0.6\textheight,keepaspectratio]{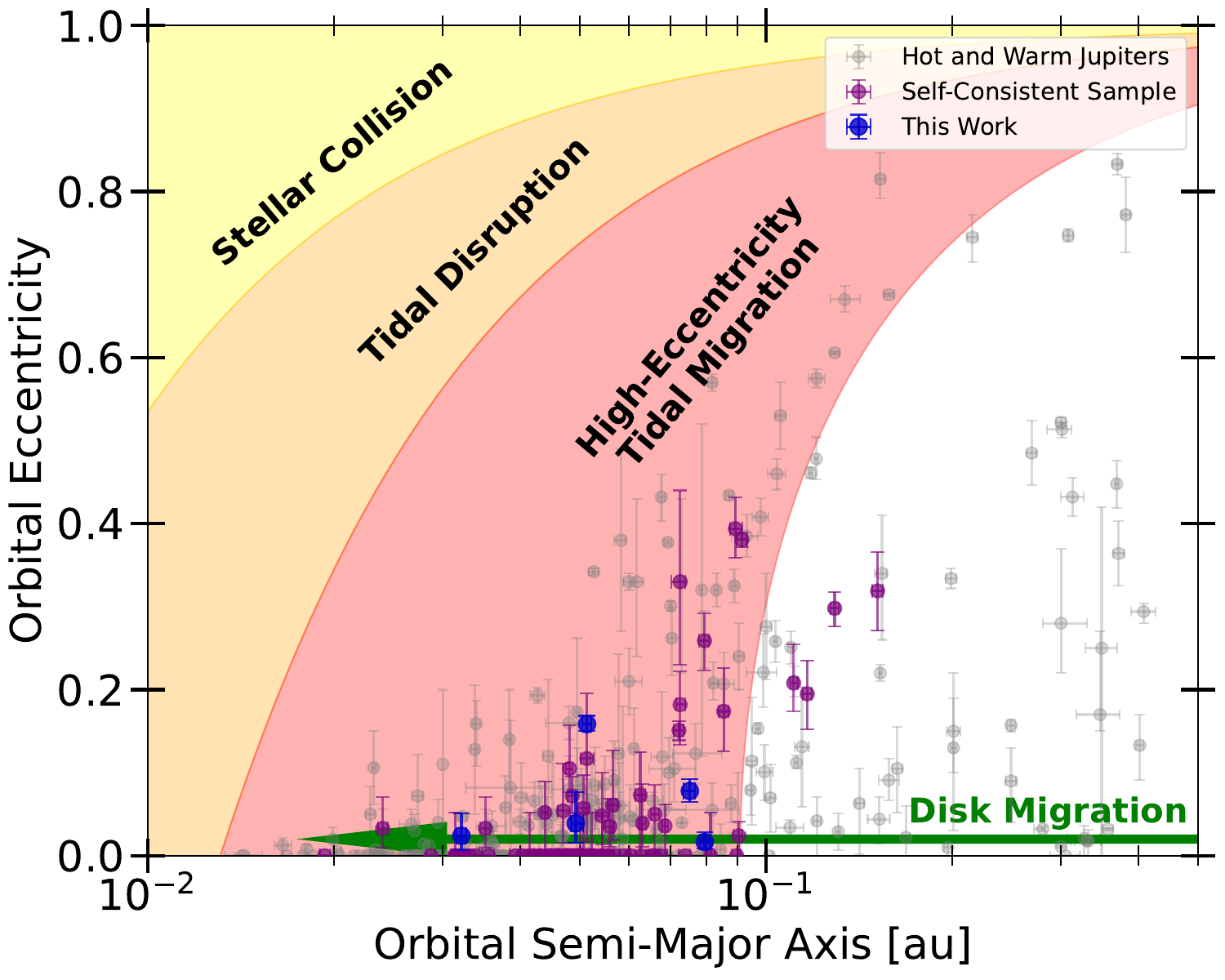}
    \caption{Eccentricity and semi-major axis distribution of the planets discovered in this work, compared to confirmed warm and hot giant planets in the literature and several avenues and outcomes of giant planet migration (adapted from Figure 4 from \citealp{Dawson:2018} and Figure 5 from \citealp{Schulte:2024}). The region labeled "Stellar Collision" corresponds to a giant planet colliding with the star, assuming the star is 1 R$_\odot$ in size. The region labeled "Tidal Disruption" corresponds to a Jupiter-like planet falling within the Roche limit of the Sun. Finally, the red region corresponds to the tidal circularization of a highly eccentric giant planet around a Sun-like star. The blue circles represent the \exofast median semi-major axis and eccentricity of the planets in this work, while the gray circles represent substellar bodies discovered by \tess with masses $> 0.25$ \mj and reported eccentricities, obtained from the NASA Exoplanet Archive on \NEAdate. Many of the eccentric planets, including those that fall outside of the three shaded regions, can be explained by planet-planet scattering that has not yet or will not be tidally circularized. The equations that describe each of these shaded regions are presented in \citealp{Dawson:2018}.}
    \label{fig:a_vs_ecc}
\end{figure*}

Among the entire population of confirmed HJs, several trends have been noted in the literature, such as the tendency for giant planets to orbit metal-rich stars \citep{Gonzalez:1997, Fischer:2005}. The five systems in this article agree with that trend; all five systems have super-solar metallicities and both TOI-5261 and TOI-6420 in particular are quite metal-rich ([Fe/H] = $0.254^{+0.076}_{-0.077}$ and $0.24^{+0.073}_{-0.077}$, respectively). \cite{Bonomo:2017}, \cite{Rodriguez:2023}, and  \cite{Zink:2023} have argued that the population of confirmed HJs has an eccentricity distribution that is consistent with high-eccentricity tidal migration being the primary mechanism for the migration of HJs. The current self-consistent sample of HJs is in agreement with this assessment, as is illustrated in Figure \ref{fig:a_vs_ecc}. Further assessment of the likelihood and frequency of this evolutionary mechanism is difficult, however, due to the incomplete nature of the sample.

\section{Summary}\label{sec:summary}

In this article, we reanalyzed the benchmark system \confirmedtoi, an inflated hot Jupiter orbiting an F-type subgiant star, and uncovered a Li absorption feature in the star's spectrum. We measured the equivalent width of the Li I doublet to be $120. \pm 13$ m\AA, which is significantly larger than a control sample of 1381 similar stars (median = 33 m\AA; median absolute deviation = 13.1 m\AA). Due to \confirmedtoi's mass and age, Li enrichment from the ingestion of 1 \mj of planetary material would have a timescale of 1.5 Gyr \citep{Soares-Furtado:2021}, implying that planetary engulfment is a likely explanation for \confirmedtoi's Li feature.

We also discovered four massive HJs orbiting FGK stars: TOI-4773 b, TOI-5261 b, TOI-5350 b, and TOI-6420 b. All four of these are more massive than 5 \mj and occupy a sparse region of the planetary mass-radius diagram, likely because they are less likely outcomes of core accretion. TOI-5261 is a solar analog that hosts an eccentric 11.49 \mj HJ which is nearly massive enough to burn deuterium in its core. This system, because of its rarity and the host star's similarity to the Sun, is a compelling target for measurements of the planet's atmosphere and stellar spin-orbit alignment to uncover additional evidence on how HJs migrate.

Each of the five systems in this paper is an important addition to the growing self-consistent catalog of HJs that have been observed by \tess and analyzed in the fashion prescribed by the MEEP survey and the TESS Grand Unified Hot Jupiter Survey \citep{Yee:2022, Yee:2023a}. This sample, which will be complete to a magnitude of at least 12.5 in \textit{Gaia}'s $G$ bandpass, will be a valuable tool for using statistics of the HJ population to infer the occurrence of HJs around different stellar types, confirm or reject tentative trends reported in the literature, and constrain the likelihood and frequency of each theory explaining the formation and evolution of HJs.

\section*{Acknowledgements}

% Jack Schulte
We thank the anonymous reviewer for their helpful comments that improved the quality of this article. Research reported in this publication was supported in part by funding provided by the National Aeronautics and Space Administration (NASA), under award number 80NSSC20M0124, Michigan Space Grant Consortium (MSGC).

% GALAH
This work made use of the Fourth Data Release of the GALAH Survey (Buder et al. 2021). The GALAH Survey is based on data acquired through the Australian Astronomical Observatory, under programs: A/2013B/13 (The GALAH pilot survey); A/2014A/25, A/2015A/19, A2017A/18 (The GALAH survey phase 1); A2018A/18 (Open clusters with HERMES); A2019A/1 (Hierarchical star formation in Ori OB1); A2019A/15, A/2020B/23, R/2022B/5, R/2023A/4, R2023B/5 (The GALAH survey phase 2); A/2015B/19, A/2016A/22, A/2016B/10, A/2017B/16, A/2018B/15 (The HERMES-TESS program); A/2015A/3, A/2015B/1, A/2015B/19, A/2016A/22, A/2016B/12, A/2017A/14, A/2020B/14 (The HERMES K2-follow-up program); R/2022B/02 and A/2023A/09 (Combining asteroseismology and spectroscopy in K2); A/2023A/8 (Resolving the chemical fingerprints of Milky Way mergers); and A/2023B/4 (s-process variations in southern globular clusters). We acknowledge the traditional owners of the land on which the AAT stands, the Gamilaraay people, and pay our respects to elders past and present. This paper includes data that has been provided by AAO Data Central (datacentral.org.au).

% SOAR
This research is also based in part on observations obtained at the Southern Astrophysical Research (SOAR) telescope, which is a joint project of the Minist\'{e}rio da Ci\^{e}ncia, Tecnologia e Inova\c{c}\~{o}es (MCTI/LNA) do Brasil, the US National Science Foundation’s NOIRLab, the University of North Carolina at Chapel Hill (UNC), and Michigan State University (MSU).

% Ivan Strackhov
The work of I.A.S. was conducted under the state assignment of Lomonosov Moscow State University.

%%%%%%%%%%%%%%%%%%%%%%%%%%%%%%%%%%%%%%%%%%%%%%%%%%
\section*{Data Availability}

The \tess\ observations used in this article, which are presented in \S \ref{subsec:TESS} and shown in Table~\ref{tab:tess} and Figures~\ref{fig:toi4138}-\ref{fig:toi6420}, are publicly available through MAST\footnote{\url{https://archive.stsci.edu/}}. The archival photometric observations from Gaia, 2MASS, and WISE, as shown in Table~\ref{tab:lit}, are available to be retrieved from VizieR\footnote{\url{https://vizier.cfa.harvard.edu/viz-bin/VizieR-2}} \citep{Ochsenbein:2000}. The follow-up time-series photometry and Speckle contrast curves which were presented in Tables \ref{tab:followup} and \ref{tab:hri} are available on ExoFOP-TESS\footnote{\url{https://exofop.ipac.caltech.edu/tess/}}. The full RV dataset used in this article, as referenced in Table~\ref{tab:rv}, is available in the online journal. Finally, the original codes that were used in the production of this paper are available at \url{https://github.com/jackschulte/MEEP2}.



\bibliographystyle{mnras}
\bibliography{main} % if your bibtex file is called example.bib

%\begin{thebibliography}{99}
%\end{thebibliography}

%%%%%%%%%%%%%%%%%%%%%%%%%%%%%%%%%%%%%%%%%%%%%%%%%%

%%%%%%%%%%%%%%%%% APPENDICES %%%%%%%%%%%%%%%%%%%%%

% \appendix

% \section{Some extra material}

% If you want to present additional material which would interrupt the flow of the main paper,
% it can be placed in an Appendix which appears after the list of references.

%%%%%%%%%%%%%%%%%%%%%%%%%%%%%%%%%%%%%%%%%%%%%%%%%%

% Don't change these lines
\bsp	% typesetting comment
\label{lastpage}
\end{document}